\documentclass[aps,prc,preprint,tightenlines,superscriptaddress,showpacs,%
amssymb,byrevtex,nofootinbib]{revtex4}
\usepackage{dcolumn}
\usepackage{epsfig}
\usepackage{bm}
\usepackage{slashed}
\usepackage[utf8]{inputenc}
\usepackage{amsmath,amssymb}
\usepackage{graphicx}
\usepackage{subfigure}
\usepackage{hyperref}
\usepackage{nicefrac}
\usepackage{xr}
\usepackage{footnote}
\usepackage{booktabs}
\usepackage{import}
\usepackage{multirow}
\usepackage{here}
\usepackage{color}
\usepackage{dsfont}
\usepackage{placeins}
\usepackage[normalem]{ulem}

\newcommand{\stkout}[1]{\ifmmode\text{\sout{\ensuremath{#1}}}\else\sout{#1}\fi}

\begin{document}
\title{Nucleon Resonances with Hidden Charm in $\gamma p $ reactions }
\author{ Jia-Jun Wu}
\affiliation{School of Physical Sciences, University of Chinese Academy of Sciences(UCAS), Beijing 100049, China }
\author{T.-S. H. Lee}
\affiliation{Physics Division, Argonne National Laboratory, Argonne, Illinois 60439, USA}
\author{Bing-Song Zou}
\affiliation{CAS Key Laboratory of Theoretical Physics, Institute of Theoretical Physics, Chinese Academy of Sciences, Beijing 100190, China  }
\begin{abstract}
The excitations of nucleon resonances with hidden charm, $N^*_{c\bar{c}}$, in the $\gamma p$  reactions
are investigated by  using the predictions from the available
meson-baryons (MB) coupled-channel models of $N^*_{c\bar{c}} $with MB = $\rho N$, $\omega N$,
$J/\psi N$, $\bar{D}\Lambda_c$, $\bar{D}^* \Lambda_c$,$\bar{D}\Sigma_c$, $\bar{D}^*\Sigma_c$, $\bar{D}\Sigma^*_c$.
For the $\gamma\,p\rightarrow J/\psi\, p$ process, we first apply
the Model of Donnachie and Landshoff to calculate the Pomeron-exchange amplitudes with the
parameters determined from
fitting the available  total cross section data up to invariant  mass $W=300$ GeV.
We then add the resonant $\gamma p \rightarrow N^*_{c\bar{c}}\rightarrow J/\psi\, p$ amplitudes to examine
the  effects of $ N^*_{c\bar{c}}$ excitations on the cross sections of $\gamma\,p\rightarrow J/\psi\, p$
in the near threshold energy region covered by the recent experiments at Jefferson Laboratory.
The $N^*_{c\bar{c}}\to MB$ transition matrix elements  are determined from the partial decay widths predicted by the
considered meson-baryons  coupled-channel models of $N^*_{c\bar{c}}$.
The $\gamma p \to N^*_{c\bar{c}}$ transition amplitudes are  calculated from the Vector Meson Dominance (VMD) model
as $\gamma p \to V p \to N^*_{c\bar{c}}$ with V = $\rho$, $\omega$, $J/\psi$.
The total $\gamma p\rightarrow J/\psi\, p$ amplitudes then depend on an off-shell form factor,
parameterized as  $F_V(q^2)=\Lambda^4/(\Lambda^4 + (q^2 - m^2_V)^2)$,
 which is needed to account for the $q^2$-dependence of the photon-vector meson coupling constant
$\frac{e\,m^2_V}{f_V}$ of the VMD model.
It has been found that with $\Lambda = 0.55$ GeV, the predicted total cross sections are within the
range of the  data in the energy region near the $J/\psi$ production threshold.
We then demonstrate that the $N^*_{c\bar{c}}$ can be most easily identified in the differential cross sections at large angles where the contribution from Pomeron-exchange becomes negligible.
With the same VMD model and the same coupled-channel models of $N^*_{c\bar{c}}$,
we also calculate the resonant amplitudes for the
 $\gamma\,p\rightarrow Vp\rightarrow  N^*_{c\bar{c}} \rightarrow \bar{D}^0\Lambda_c^+ (\bar{D}^{*0}\Lambda_c^+)$
processes.
 By adding the non-resonant amplitudes
due to the exchange of $\bar{D}^{*0}$ ($\bar{D}^{0})$,
we then predict the cross sections of $\gamma\,p\rightarrow \bar{D}^0\Lambda_c^+
 (\bar{D}^{*0}\Lambda^+_c)$ for additional  experimental tests of the available meson-baryon coupled-channel
models of $N^*_{c\bar{c}}$.

\end{abstract}

\maketitle

\section{Introduction}

It is well recognized that
the interaction between the nucleon ($N$) and a $c\bar{c}$ system of charm quark($c$) and anti-charm quark($\bar{c}$) is
mainly due to the gluon-exchange mechanisms.
All of the earlier
investigations~\cite{Peskin:1979va, Bhanot:1979vb, Luke:1992tm, Brodsky:1997gh, Kawanai:2010ev, Kaidalov:1992hd, Donnachie:1984xq,  Brodsky:1989jd}
have indicated that the $c\bar{c}$-N
interaction is attractive.
This implies the possible existence of
nuclear systems with hidden charm, as investigated in Refs.\cite{Brodsky:1989jd,Wu:2012wta,Wu:2013xma}.
For the baryon number $B=1$ system, it was proposed~\cite{Wu:2010jy} in 2010 that
there exists excited nucleons with $c\bar{c}$ components in the mass range of 4.0 - 5.0 GeV
within a  meson-baryon coupled-channel  model.
Such baryons with hidden charm  were subsequently also predicted~\cite{Wu:2010vk,Yang:2011wz,Wu:2012md, Oset:2012ap,
Garcia-Recio:2013gaa,Xiao:2013yca,Uchino:2015uha} as molecular states
made of anti-charmed mesons and charmed baryons (such as $\bar{D}^{(*)}\Sigma^{(*)}_c$).
Alternatively, they are
described as compact pentaquark states made of colored quark clusters~\cite{Yuan:2012wz} or a mixture of the two configurations~\cite{Wang:2011rga} .
The masses from these  earlier predictions are qualitatively consistent with  the
mass ($m$) and width ($\Gamma$) of two Pentaquark states ($P_c$) identified from
analyzing the $J/\psi$-$p$ invariant mass distributions
of the  $\Lambda^*_c\rightarrow K J/\psi p$ decays measured
 by the LHCb collaboration~\cite{Aaij:2015tga, Aaij:2016ymb} in 2015.
Their results
are listed in the left part of Table \ref{tab:LHCbdata}.
\begin{table}[ht]
\setlength{\tabcolsep}{0.15cm}

\begin{center}\caption{The masses ($m$(MeV)) and total widths ($\Gamma$ (MeV) of $P_c$ reported in Ref.~\cite{Aaij:2015tga, Aaij:2016ymb, Aaij:2019vzc}}

\begin{tabular}{ccc|ccc}\hline


& 2015 &

& & 2019 & \\

$P_c$ & $m$ & $\Gamma$

& $P_c$ & $m$ & $\Gamma$ \\

\hline

& &

& $P_c(4312)$ & $4311.9\pm0.7^{+6.8}_{-0.6}$ & $9.8\pm2.7^{+3.7}_{-4.5}$ \\

$P_c(4380)$ & $4380\pm8\pm 29$ & $205\pm18\pm 86$

& & & \\

$P_c(4450)$ & $4449.8\pm1.7\pm2.5$ & $39\pm5\pm 19$

& $P_c(4440)$ & $4440.3\pm1.3^{+4.1}_{-4.7}$ & $20.6\pm4.9^{+8.7}_{-10.1}$ \\

& &

& $P_c(4457)$ & $4457.3\pm0.6^{+4.1}_{-1.7}$ & $6.4\pm2.0^{+5.7}_{-1.9}$ \\

\hline

\end{tabular} \label{tab:LHCbdata}

\end{center}

\end{table}

The resonance peaks in the $J/\psi$-$p$ invariant mass distributions from the
LHCb measurement~\cite{Aaij:2015tga}
had motivated a lot of theoretical
efforts~\cite{Chen:2015loa, Chen:2015moa, Roca:2015dva, He:2015cea, Huang:2015uda, Wang:2015qlf, Yang:2015bmv, Chen:2016heh, Roca:2016tdh, Lu:2016nnt, Shimizu:2016rrd, Shen:2016tzq, Ortega:2016syt,Meissner:2015mza, Yamaguchi:2016ote, He:2016pfa, Oset:2016nvf, Xiao:2016ogq, Lin:2017mtz, Yamaguchi:2017zmn, Shen:2017ayv, Lin:2018kcc, Shimizu:2019jfy, Anwar:2018bpu, Eides:2018lqg, Burns:2015dwa,Takeuchi:2016ejt, Xiang:2017byz, Li:2017kzj, Hiyama:2018ukv, Chen:2016otp, Lebed:2015tna, Li:2015gta,Wang:2015epa, Zhu:2015bba, Guo:2015umn, Liu:2015fea, Guo:2016bkl, Bayar:2016ftu,Chen:2016qju, Zhao:2016akg, Dong:2017gaw, Guo:2017jvc,Ali:2017jda}.
Roughly speaking, there are three different interpretations of these peaks:
\begin{enumerate}
\item they are due to the excitations of meson-baryon   molecular systems  which could be made of   : (1)
 anti-charm  mesons and charm baryons~\cite{Chen:2015loa, Chen:2015moa, Roca:2015dva, He:2015cea, Huang:2015uda, Wang:2015qlf, Yang:2015bmv, Chen:2016heh, Roca:2016tdh, Lu:2016nnt, Shimizu:2016rrd, Shen:2016tzq, Ortega:2016syt,Meissner:2015mza, Yamaguchi:2016ote, He:2016pfa, Oset:2016nvf, Xiao:2016ogq, Lin:2017mtz, Yamaguchi:2017zmn, Shen:2017ayv, Lin:2018kcc, Shimizu:2019jfy},
 (2) baryons and charmonium~\cite{Anwar:2018bpu, Eides:2018lqg},
(3) the mixture~\cite{Burns:2015dwa,Takeuchi:2016ejt} of (1) and (2).
\item they could be the multi-quark states within the conventional
constituent quark model~\cite{Xiang:2017byz, Li:2017kzj, Hiyama:2018ukv},
or the cluster states pictured as a diquark-diquark-antiquark system~\cite{Chen:2016otp, Lebed:2015tna, Li:2015gta,Wang:2015epa} or a diquark-triquark system~\cite{Zhu:2015bba}.
\item  $P_c(4450)$ may not be a resonance state because it is close to the
 triangle singularity~\cite{Guo:2015umn, Liu:2015fea, Guo:2016bkl} and the observed narrow peak is purely due to kinematic effect, although for some quantum numbers of $P_c$ state preferred in Ref.\cite{Aaij:2015tga}, such as $ 3/2^-$ or $5/2^+$, the TS can not explain the peak as shown in Ref.~\cite{Bayar:2016ftu}.

\end{enumerate}

With the the new results from the  LHCb collaboration~\cite{Aaij:2019vzc},
these theoretical interpretations can be better tested.
By analyzing the data which are about a factor of 9 more than what they analyzed in 2015, LHCb collaboration obtained three clean peaks which are interpreted as the excitations of  three Pentaquark states, as listed in the right part of Table \ref{tab:LHCbdata}.
Comparing with their results of 2015, the main features of these new data are:
(1) $ P_c(4312)$  could be a new Pentaquark state near $\Sigma_c \bar{D}$ threshold.
(2) $P_c(4440)$ and $P_c(4457)$  were two narrow states which could not be resolved in their 2015 determination of $P_c(4450)$.
(3) $P_c(4380)$ with about 200 MeV width of 2015 could be the very broad state and is not given mass and width in this analysis.
It is important to note that these three narrow states are all just below the corresponding anti-charmed meson-charmed baryon threshold and hence the simplest interpretation is that  they are made of meson-baryon components, as suggested
in Refs.\cite{Chen:2019asm, Chen:2019bip, Liu:2019tjn, Guo:2019fdo, He:2019ify, Liu:2019zoy, Huang:2019jlf, Xiao:2019mvs, Shimizu:2019ptd, Guo:2019kdc,Ali:2019npk}.
%

%
The nucleon resonances with hidden charm, called $N^*_{c\bar{c}}$ from now on in this paper,
can also be investigated by using the electromagnetic production of $J/\psi$ from the nucleon, such as
 $e\,+p\rightarrow e'\,+ J/\psi\,+p$ studied in Refs.\cite{Wu:2010vk, Wu:2010jy}.
The prediction of $\gamma p \to N^*_{c\bar{c}}\to J/\psi p$ cross section within the coupled-channel model
of Ref.~\cite{Wu:2010vk}  was then made in Ref.\cite{Huang:2013mua} by using the Vector Meson Dominance (VMD) Model to generate vector (V) mesons, $\rho$, $\omega$, and $J/\psi$, from photon.
%
%
Few more predictions of $\gamma p \to N^*_{c\bar{c}}\to J/\psi p$ had been
made~\cite{Wang:2015jsa,Kubarovsky:2015aaa,Karliner:2015voa,Blin:2016dlf, Paryev:2018fyv,Wang:2019krd}
within the meson-baryon coupled-channel model since 2015.
The differences between these works are in their choice of $N^*_{c\bar{c}}$ model,
vector mesons included in using VMD, and
the background amplitudes which could be calculated from Pomeron-exchange or
2-gluons and 3-gluons exchange model.
%
%
%
%
%
%

%
%

In parallel to these theoretical efforts, 
an experiment~\cite{Meziani:2016lhg, Hafidi:2017bsg}(JLab(E12-16-007)) on
$\gamma\, p \rightarrow J/\psi\, p$ near threshold at JLab(E12-16-007) was approved in 2016 and the data from this effort will soon become available.
A separate effort at JLab using GlueX detector \cite{Dobbs:2017vjw} has recently
published~\cite{Ali:2019lzf}
their measured total cross sections of $\gamma p \rightarrow J/\psi\,p$.
 The main purpose of this work is to provide information for
examining whether the $N^*_{c\bar{c}}$ predicted by the available meson-baryon coupled-channel models
can be observed in the  data from these two experiments.

To proceed, it is necessary to first recognize that $P_c$ states reported by the LHCb
collaboration are from the measurements of $J/\psi$-$p$ invariant mass distribution of the
$\Lambda_c \to K\,+ J/\psi\,+ p$ decay.
Thus the information one can use to test the available $N^*_{c\bar{c}}$ models is the total widths and  masses of the reported $P_c$ states.
The spins and parities of these states can not be determined since a partial-wave analysis of
 $\Lambda_c \to K\,+ J/\psi\,+ p$ decays requires detailed angular distribution data, not just the
invariant mass distributions.
Accordingly, one can not determine the partial decay width for each possible meson-baryon
channels of $N^*_{c\bar{c}}\rightarrow J/\psi N, \rho N, \omega N, \bar{D}\Lambda_c, \bar{D}^*\Lambda_c, \bar{D}\Sigma_c, \bar{D}^*\Sigma_c...$.
Here we also mention that two of the resonance peaks reported by LHCb collaboration
are near the threshold of $\bar{D}^{(*)}\Sigma_c$ channel and thus the
identification of resonances in this region
 must account for the cusp effect in a analysis constrained by the three-body unitarity.
The importance of three-body unitarity in analyzing
the three-body decays of heavy mesons have been
demonstrated~\cite{Kamano:2011ih} recently, but is not considered in the analysis of
LHCb collaboration.
Therefore,
no attempt will be made here to revise the considered meson-baryon models
to reproduce the resonance peaks of the LHCb data.
Instead,  we will only
consider the available models which have predicted  $N^*_{c\bar{c}}$ with masses within the range of the LHCb data.
By using the  spins, parities, and partial decay widths from those models,
we can then use the VMD to predict the amplitudes of
$\gamma\, p\rightarrow N^*_{c\bar{c}}\rightarrow J/\psi\,\,p$.
Here we notice that
the VMD coupling constant $g_{\gamma,V}=e\,m^2_V/f_V$
for the $\gamma \rightarrow V$ transitions for $V=\rho, \omega, J/\psi$  are
 conventionally determined from the decay widths of $V\rightarrow \gamma \rightarrow e^+e^-$ with $q^2=m^2_V$ of the intermediate $\gamma$.
In the situation of $\gamma p \rightarrow Vp \rightarrow N^*_{c\bar{c}}$, we have $q^2=0$,
i.e. the intermediate vector is far off-mass-shell, and
thus the VMD parameter $g_{\gamma,V}$ must be modified to account for this  $q^2$-dependence.
%
%
Ideally, this $q^2$-dependence should be calculated from a QCD model as done in Ref.\cite{Pichowsky:1996tn}.
Here we will treat it as a phenomenological part of our calculation by introducing a off-shell form factor
$F_V=\Lambda^4/(\Lambda^4+(q^2,m^2_V)^2)$ with $\Lambda$ determined by the available total cross section
data, as will be explained in section III.
We also make sure that the parametrization of VMD is gauge invariant when the off-shell form factor is included.

To predict the cross sections of $\gamma p\rightarrow J/\psi p$ ,
it is necessary to include the non-resonant amplitudes due to the gluon-exchange mechanisms.
In this work, we use the model of Donnachie and Landshoff~\cite{Donnachie:1984xq}
within which the gluon-exchange mechanism is  phenomenologically
parametrized as Pomeron-exchange within the Reggy Phenomenology of high energy reactions.
By fitting the total cross section data up to very high energy $W=300$ GeV, the Pomeron parameters are well determined and can be used to define the non-resonant amplitudes
in the near threshold region of our interest in this paper. Our approach is therefore different from
the approaches using the models
of two-gluon and three-gluon exchange of Refs.\cite{Brodsky:2000zc}
, as will be discussed
 later.

For  additional studies of $N^*_{c\bar{c}}$ excitations,  
 we have also explored other meson photo-production processes which do not have
Pomeron-exchange mechanisms. We have  found that  experiments on 
 $\gamma\,p\rightarrow \rightarrow  \bar{D}^0\Lambda_c^+ (\bar{D}^{*0}\Lambda_c^+)$
could be useful.
With the same VMD model and the same coupled-channel models of $N^*_{c\bar{c}}$,
we have calculated the resonant amplitudes for the
 $\gamma\,p\rightarrow Vp\rightarrow  N^*_{c\bar{c}} \rightarrow \bar{D}^0\Lambda_c^+ (\bar{D}^{*0}\Lambda_c^+)$
processes.
 By adding the non-resonant amplitudes
due to the exchange of $\bar{D}^{*0}$ ($\bar{D}^{0})$,
we then predict the cross sections of $\gamma\,p\rightarrow \bar{D}^0\Lambda_c^+
 (\bar{D}^{*0}\Lambda^+_c)$ for additional  experimental tests of the available meson-baryon coupled-channel
models of $N^*_{c\bar{c}}$.

The paper is organized as follows.
In section II, we give formulas for calculating the cross sections of $\gamma\,\,p\rightarrow J/\psi\,\,p$,
and present  formula for calculating the Pomeron-exchange amplitude and the
$\gamma \,\,p \rightarrow N^*_{c\bar{c}}\rightarrow J/\psi\,\, p$ resonant amplitudes.
In section III, we present our results for $\gamma\,\,p\rightarrow J/\psi\,\,p$.
The results for $\gamma\,\,p\rightarrow \bar{D}^0\Lambda_c, \bar{D}^{*0}\Lambda_c$ are given in section IV.
The discussion and summary will be given in the last section.

\section{Cross section formula for vector meson photo-production reaction on the nucleon}

We consider the photo-production of a Meson ($M$)-Baryon ($B$) system :
 $\gamma(q) + p (p) \rightarrow M(q') + B(p')$.
In the center of mass system, the four-momentum of these particles can be defined as
\begin{eqnarray}
q&=&(k,\,\vec{k})
\nonumber\\
q'&=&(E_M(k'),\vec{k}')
\nonumber\\
p&=&(E_N(k),- \vec{k})
\nonumber\\
p'&=&(E_B(k),- \vec{k}')
\nonumber\\
P&=&q+q'=p+p'=(W,\vec{0}),
\nonumber
\end{eqnarray}
where
$k$($k'$) is the length of three momenta $\vec{k}$($\vec{k}'$),
$E_a(k)=\sqrt{m^2_a+k^{2}}$ is the energy of a particle with
mass $m_a$, and $W$ is the invariant mass of system.
For a given $W$ and angle ($\Omega$) between $\vec{k}$ and $\vec{k}'$,
all of the above kinematic variables are determined  by $W=k+E_N(k) = E_M(k')+E_B(k')$.
The differential cross section  can then  be written
\begin{eqnarray}
\frac{d\sigma}{d\Omega}
&=&
\frac{(2\pi)^4}{k^2}\rho_{\gamma N}(k)\rho_{MB}(k')
\frac{1}{4}\sum_{\lambda_\gamma,\lambda_{M}}\sum_{m_s,m'_s}
|<\vec{k}'\lambda'_{M}m'_s|T(W)|\vec{k}\lambda_\gamma m_s>|^2 \,,
\label{eq:crst}
\end{eqnarray}
where $\lambda'_{M}$ ($\lambda_\gamma$) is the helicity of
the meson $M$ (photon $\gamma$), $m_s$( $m'_s$) is the z-component of the
spin of initial proton $p$ (final baryon  $B$).
 The phase space factors in Eq.(\ref{eq:crst}) are
\begin{eqnarray}
\rho_{\gamma N}(k)&=&\frac{k^2E_N(k)}{W} \nonumber \\
\rho_{MB}(k')&=&\frac{k'E_M(k')E_N(k')}{W}
\end{eqnarray}



The reaction amplitude is written as
\begin{eqnarray}
<\vec{k}'\lambda'_{M}m'_s|T(W)|\vec{k}\lambda_\gamma m_s>
& =&
\frac{1}{(2\pi)^3}\sqrt{\frac{ m_Nm_B }{4 k E_N(k)E_{M}(k')E_B(k') }}
\epsilon_\nu(q,\lambda_\gamma)
[j^{\nu}_{\lambda'_{M},\,m'_s,\,m_s}(q',p',q,p)]
\nonumber\\
\label{eq:tot-amp}
\end{eqnarray}
where $\epsilon_\nu(q,\lambda_\gamma)$ is
the polarization vector of photon,
and $j^{\nu}_{\lambda'_{M},\,m'_s,\,m_s}(q,p,q',p')$ is a Lorentz covariant current
matrix element.
For the vector meson photo-production $\gamma p \to J/\psi p $ ($M=J/\psi$ and $B=N$) process,
the current matrix element can be written
as
\begin{eqnarray}
j^{\nu}_{\lambda'_{J/\psi},\,m'_s,\,m_s}(q,p,q',p')=\bar{u}_{p}(p',m'_s)
\epsilon^*_\mu(q',\lambda'_{J/\psi})\mathcal{M}^{\mu\nu}(q,p,q',p')
u_{p}(p,m_s)\,, \label{eq:ampjpsip}
\end{eqnarray}
where $u_{a}(p,m_s)$ is the spinor of the baryon $a$  (with the normalization
$\bar{u}_{a}(p,m_s){u}_{a}(p,m'_s) = \delta_{m_s,m'_s}$) ,
$\epsilon_\nu(q,\lambda'_J/\psi)$ is the polarization vector of $J/\psi$.
The current matrix element must satisfy
the gauge invariance condition $j^{\nu}q_{\nu} = 0$.

In this work, we assume
 that the $J/\psi$  photo-production
amplitudes $\mathcal{M}^{\mu\nu}(q,p,q',p')$ of Eq.(\ref{eq:ampjpsip})
can be written as
\begin{eqnarray}
\mathcal{M}^{\mu\nu}(q,p,q',p') =
\mathcal{M}_{\mathbb{P}}^{\mu\nu}(q,p,q',p')+\mathcal{M}_{N^*}^{\mu\nu}(q,p,q',p')
\label{eq:amp}
\end{eqnarray}
where $\mathcal{M}_{\mathbb{P}}^{\mu\nu}(q,p,q',p')$ is the Pomeron-exchange amplitude
of Donnachie and Landshoff,
and $\mathcal{M}_{N^*}^{\mu\nu}(q,p,q',p')$ is the
$\gamma N \rightarrow N^*_{c\bar{c}}\rightarrow J/\psi\,\,N$ amplitude.
In the following, we will describe the calculations of these two amplitudes.

\subsection{Pomeron-exchange mechanism}

\begin{figure}[h]
\centering
\includegraphics[width=0.25\columnwidth,angle=-90]{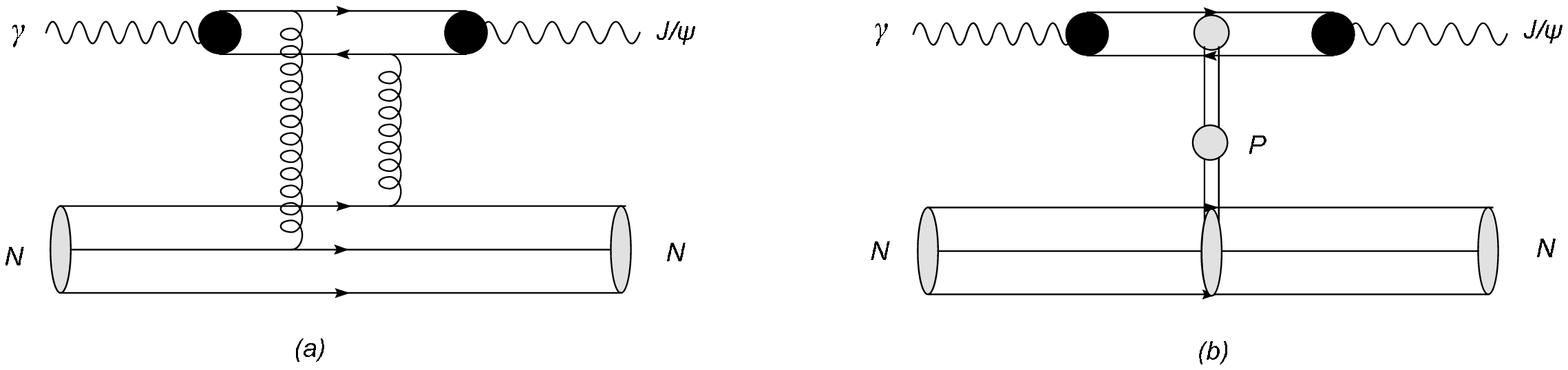}
\caption{
Gluon-exchange mechanism
of $\gamma N \rightarrow J/\psi +N$}
\label{fig:pom-mec}
\end{figure}

It is well recognized that the photo-production of $J/\psi$ from the nucleon  is mainly due to
gluon-exchange mechanism, such as the leading
two-gluon exchange mechanism illustrated in  Fig.\ref{fig:pom-mec} (a).
It is also known that Pomeron-exchange has been an essential element in Reggy Phenomenology.
Within the model of Donnachie and Landshoff (DL) \cite{Donnachie:1984xq},
it is assumed that Pomeron ($\mathbb{P}$) can be identified with gluons and
the Pomeron-exchange mechanism can be parametrized in terms of Pomeron-quark coupling
constant $\beta_q$ and appropriately form factors at the
$\mathbb{P}\,J/\psi\rightarrow J/\psi$  and $\mathbb{P}\,N\rightarrow N$ vertices.
The DL model is illustrated in Fig.\ref{fig:pom-mec} (b).
%
%
Following a study of non perturbative two-gluon exchanges~\cite{Landshoff:1986yj},
they further assume the Pomeron-Photon analogy that the Pomeron
 can be treated as a $C=+1$ isoscalar photon to parametrize the quark-Pomeron vertex.
Thus the $\mathbb{P}\,N\rightarrow N$ vertex  can be expressed in term of
the isoscalar electromagnetic form factor of the nucleon.
Following Ref.~\cite{Oh:2002rb},
the Pomeron-exchange amplitude in Eq.(\ref{eq:amp})
is  written as:
\begin{equation}
\mathcal{M}^{\mu\nu}_\mathbb{P}(q,p,q',p') = G_\mathbb{P}(s,t)
\mathcal{T}^{\mu\nu}_\mathbb{P}(q,p,q',p')
\label{eq:MP}
\end{equation}
with
\begin{eqnarray}
\mathcal{T}^{\mu\nu}_\mathbb{P}(q,p,q',p') = i 12 \frac{eM_V^2}{f_V}
\beta_{q_{V}}F_V(t)\beta_{u/d} F_1(t)\left[ q\!\!\!/ \, g^{\mu\nu} - q^\mu \gamma^\nu \right]  \, ,
\label{eq:pom-a}
\end{eqnarray}
where $\beta_{q_{V}}$ ($\beta_{u/d}$) defines the coupling of the Pomeron with the
quark $q_{V}$ ($u$ or $d$ )in the vector meson $V$ (nucleon $N$).
Here we have introduced
the form factor for the Pomeron-vector meson vertex  as
\begin{eqnarray}
F_V(t)=\frac{1}{M_V^2-t} \left( \frac{2\mu_0^2}{2\mu_0^2 + M_V^2 - t} \right)
\label{eq:f1v}
\end{eqnarray}
where $t=(p-p')^2$. By using the Pomeron-photon analogy mentioned above,
the form factor for the Pomeron-nucleon vertex is defined by
the isoscalar electromagnetic form factor of the nucleon as
\begin{equation}
F_1(t) = \frac{4M_N^2 - 2.8 t}{(4M_N^2 - t)(1-t/0.71)^2}.
\label{eq:f1}
\end{equation}
Here $t$ is in unit of GeV$^2$, and $M_N$ is the proton mass.
Note that the factor $ \frac{eM_V^2}{f_V}$ in Eq.(\ref{eq:pom-a}) implies a  relation between the DL model and
the VMD.

\begin{figure}[t] \vspace{-0.cm}
\begin{center}
\includegraphics[width=1.0\columnwidth,angle=0]{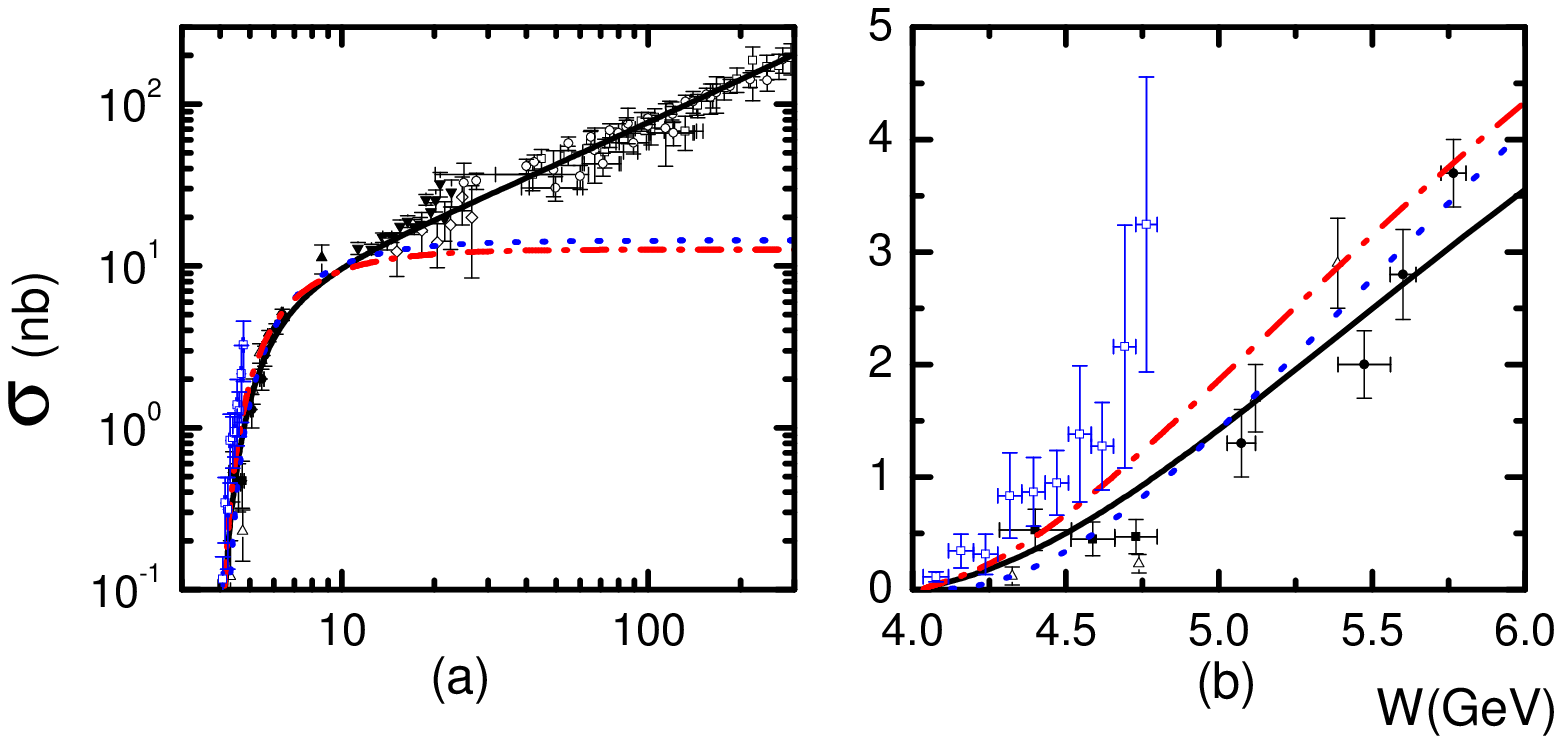}
\caption{Total cross sections of photo-production of $J/\psi$  on the proton target.
The solid curve is from the DL model of Pomeron-exchange.The dotted (dash-dotted) curves are from
the 2-gluon exchange ( 2-gluon+3 gluon-exchange) models.
The experimental data can be found in Ref.~\cite{Wu:2012wta}, except blue open spares are from Ref.~\cite{Dobbs:2017vjw}}
\label{fg:totcrst-all-v}
\end{center}
\end{figure}

\begin{figure}[htbp] \vspace{-0.cm}
\begin{center}
\includegraphics[width=0.6\columnwidth,angle=0]{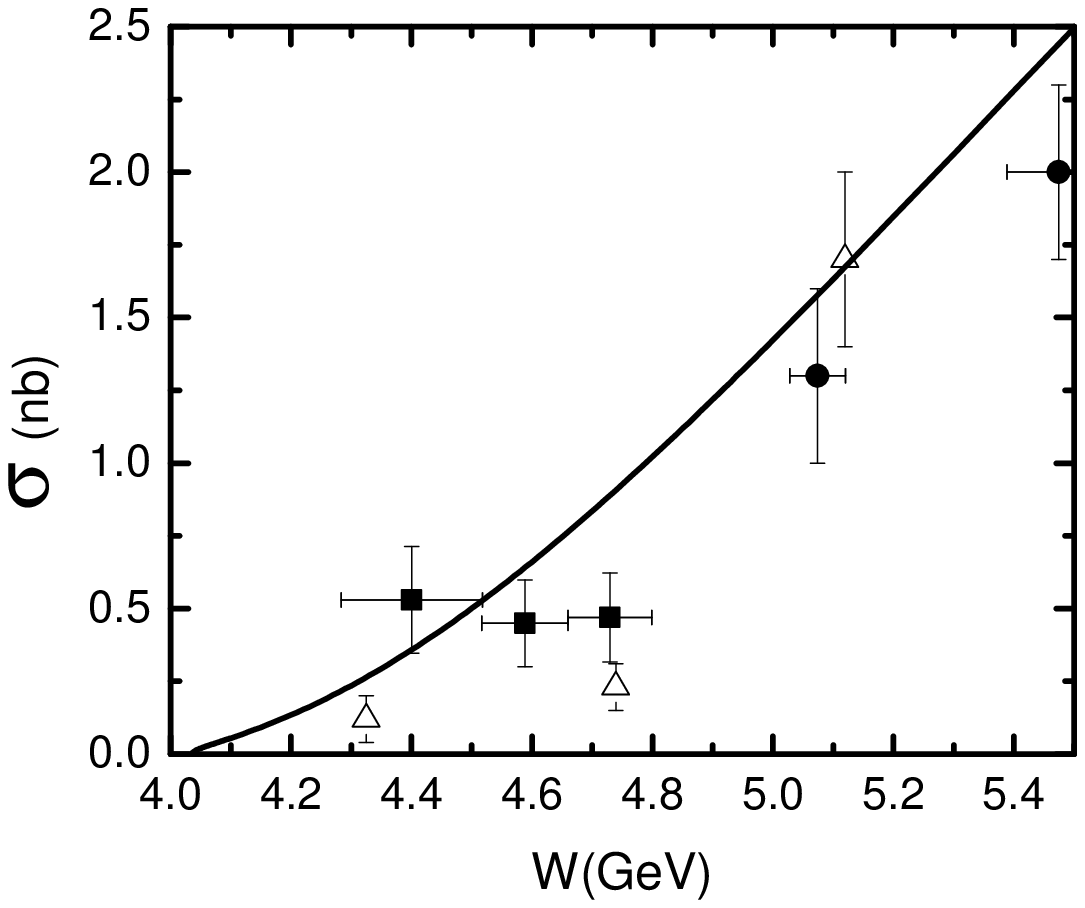}

\caption{The total cross section of $\gamma p \to J/\psi + p$ via Pomeron-exchange with the center mass energy $W$ in the near threshold energy region.
The experimental data are from Ref.~\cite{Gittelman:1975ix,Anderson:1976sd, Camerini:1975cy}
}
\label{fg:backg}
\end{center}
\end{figure}

The crucial ingredient of the Reggy Phenomenology is the propagator $G_\mathbb{P}$
for the Pomeron in Eq.(\ref{eq:MP}).
It is of the following form :
\begin{equation}
G_\mathbb{P} = \left(\frac{s}{s_0}\right)^{\alpha_P(t)-1}
\exp\left\{ - \frac{i\pi}{2} \left[ \alpha_P(t)-1 \right]
\right\} \,,
\label{eq:regge-g}
\end{equation}
where $s=(q+p)^2=W^2$, $ \alpha_P (t) = \alpha_0 + \alpha'_P t$.
By fitting the data of $\rho^0$, $\omega$, and $\phi$ photo-production\cite{Oh:2002rb}, the parameters of the model have been determined:
 $\mu_0=  1.1$ GeV$^2$, $\beta_{u/d}=2.07$ GeV$^{-1}$,
$\beta_{s}=1.6$ GeV$^{-1}$, $\alpha_0=1.08$ and $\alpha'_P = 1/s_0 = 0.25$ GeV$^{-2}$.
In our previous paper~\cite{Wu:2012wta}, we found that with the same $\mu_0^2$, $\beta_{u/d}$, and $\alpha'_P $, the $J/\psi$ photo-production  data can be fitted by setting
$\beta_c = 0.84$ GeV$^{-1}$ and choosing a larger $\alpha_0=1.25$.
In  the left side of Fig.\ref{fg:totcrst-all-v}, the results (black solid curves) from the constructed Pomeron-exchange model
are compared with all of
the  the total cross section data
of $\gamma\, p \rightarrow  J/\psi\, p$
up to invariant mass $W=$ 300 GeV.
Here we  note that the
two-gluon (dotted curves) and three-gluon (dot-dashed curves)
exchange models, with the parameters given in Refs.\cite{Brodsky:2000zc}
can not describe the data above  about $W= 10$ GeV.
The new data from JLab GlueX collaboration
 are considerably larger in magnitudes than the previous data,
as can be seen
more clearly in the right side of Fig.\ref{fg:totcrst-all-v}.
While these  data can be better described by the 2g+3g exchange model, as also shown by
the GlueX collaboration, they need further confirmation from  separate experiments at JLab.
Thus our study of $N^*_{c\bar{c}}$  starts with Fig.\ref{fg:backg} in which  the
data before 2018 are compared with  the results calculated from using
 the  Pomeron-exchange model.

%
%

%
%
%


\subsection{Excitation of $N^*_{c\bar{c}}$ resonances}

We focus on the $N^*_{c\bar{c}}$ predicted by the meson-baryon coupled-channel models with the parameters constrained by the SU(4) symmetry and the fit to the meson-baryon reaction data.
Alternatively, $N^*_{c\bar{c}}$ can be predicted by constituent quark models or non-perturbative QCD models.
These are however not considered in this work.

\begin{figure}[t] \vspace{-0.cm}
\begin{center}
\includegraphics[width=0.3\columnwidth,angle=270]{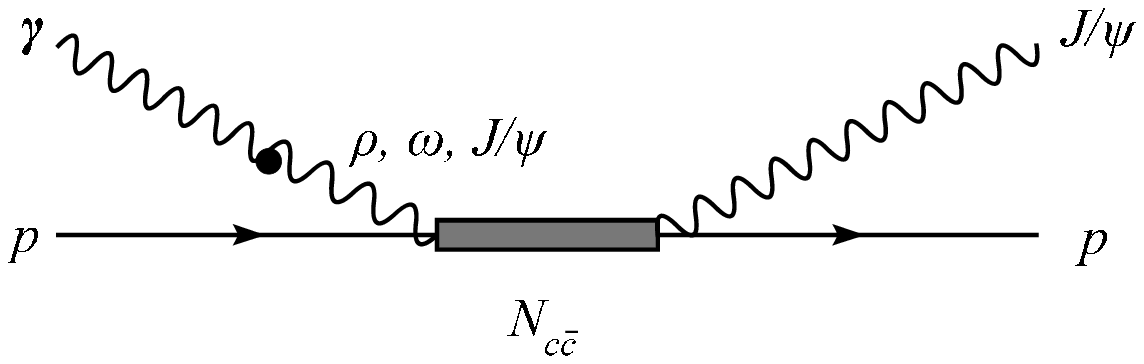}
\caption{The diagram for $\gamma p \to N_{c\bar{c}} \to J/\psi p $ with the VMD by $\rho$, $\omega$ and $J/\psi$ coupled with $\gamma$ .
}
\label{fg:ncc}
\end{center}
\end{figure}

In Table \ref{tab:mwb}, we list the predictions from most, if not all, of the coupled-channel models of $N^*_{c\bar{c}}$ in the literatures.
The relative importance of the predicted $N^*_{c\bar{c}}$ in determining $\gamma\,\, p \rightarrow J/\psi\,\, p$
 can be estimated by using a well known relation between the total cross section
$\sigma^{(tot)}$ at resonance energy $W=M_R$ and the partial decay widths
$\Gamma_{N^*_{c\bar{c}},J/\psi\,p}$ of $N^*_{c\bar{c}}\rightarrow J/\psi\,\,p$,
 $\Gamma_{N^*_{c\bar{c}},\gamma\,p}$ of $N^*_{c\bar{c}}\rightarrow \gamma \,\,p$, and the
total width $\Gamma^{(tot)}_{N^*,c\bar{c}}$:
\begin{eqnarray}
\sigma^{(tot)}(W=M_R)=\frac{2J+1}{4}\frac{4\pi}{q^2_R}
 \frac{\Gamma_{N^*_{c\bar{c}},J/\psi\,p}\Gamma_{N^*_{c\bar{c}},\gamma\,p}}
{[\Gamma^{(tot)}_{N^*_{c\bar{c}}}]^2}
\label{eq:R-dy}
\end{eqnarray}
where $J$ is the spin of $N^*_{c\bar{c}}$, and
 $q_R$ is defined by the resonance mass by $M_R=\sqrt{M^2_N+q^2_R}+q_R$.
We note here that except the model by Lin et al\cite{Lin:2018kcc} the decay width
$\Gamma_{N^*_{c\bar{c}},\gamma\,p}$ to the $\gamma p$ channel are not predicted by the models listed in Table \ref{tab:mwb}.
Thus the only way we can use these models is to use the VMD model
to describe the excitation of $N^*_{c\bar{c}}$ as
the $\gamma\,\, N \rightarrow V\,\,N \rightarrow N^*_{c\bar{c}}$ mechanism with $V=\rho,\omega J/\psi$, as
 illustrated in Fig.\ref{fg:ncc}.

In Table \ref{tab:mwb}, we also see that the predicted $N^*_{c\bar{c}}$ mainly decay into channels associated with the $\bar{D}$ meson and charmed $\Sigma_c$ baryons, as specified as "Main Channel" in the table.
However, the available energy at JLab is not high enough to investigate
the $ \gamma\,p\rightarrow \bar{D}\Sigma_c$ process.
Instead the experiment on the process $\gamma\,p\rightarrow \bar{D}\Lambda_c$ may be possible.
Thus we will also consider the $\gamma\, p \rightarrow \bar{D}\,\Lambda_c$ reaction which does not have Pomeron-exchange mechanism.
This can be studied using the models which also provide partial decay widths of $N^*_{c\bar{c}}\rightarrow \bar{D}\Lambda_c$, as also shown in Table \ref{tab:mwb}.

To proceed, we recall that the VMD is defined  by the following Lagrangian:
\begin{eqnarray}
L_{VMD}(x)=\frac{em^2_V}{f_V}A_\mu(x)\phi_V^\mu(x)
\label{eq:l-vmd}
\end{eqnarray}
where $ m_V$ is the mass of the vector meson $V$, $A_\mu$ and $ \phi_V^\mu$ are the
field operators for the photon and vector meson, respectively. The width of $V\rightarrow e^+e^-$ can then
be calculated by
\begin{eqnarray}
\Gamma_{V\rightarrow e^+e^-} =\frac{1}{3}\alpha^2m_V\frac{4\pi}{f^2_V}
\end{eqnarray}
By using the data of $\Gamma_{V\rightarrow e^+e^-}$, the decay constants of Eq.(\ref{eq:l-vmd})
can be  determined :
$f_\rho = 5.33$, $f_\omega = 15.2$, $f_\phi = 13.4$, and $f_{J/\psi} = 11.2$.
For our later discussions, we here note that these coupling constants are determined at the photon
four-momentum $q^2= m^2_V$. Thus the use of the Lagrangian Eq.(\ref{eq:l-vmd}) in other
processes with real photon $q^2=0$, a model must be used to account for the off-shell effects on these coupling
constant. In our calculations, we thus will set
\begin{eqnarray}
&&\frac{1}{f_V}\rightarrow \frac{1}{f_V} F_V(q^2)
\label{eq:off-shell} \\
&& F_V(m^2_V)=1
\label{eq:off-shell2}
\end{eqnarray}
Ideally $F_V(q^2)$ should be calculated from  the quark-loop mechanism
$V\rightarrow q\bar{q}\rightarrow \gamma (q^2)$ within a non-perturbative QCD model.
Here, we  will  determine it phenomenologically, as will be specified later.

With VMD, the $\gamma N \rightarrow N^*_{c\bar{c}}$ amplitude  can be calculated from
$\gamma N \rightarrow V_{VMD}N \rightarrow  N^*_{c\bar{c}}$, where $V_{VMD}=\rho,\omega, J/\psi$ and
 $V_{VMD}N \rightarrow  N^*_{c\bar{c}}$
calculated from
the considered meson-baryon coupled-channel models of $N^*_{c\bar{c}}$. The full amplitude
$\gamma N \rightarrow N^*_{c\bar{c}}\rightarrow VN$ can then calculated from by
using  $N^*_{c\bar{c}}\rightarrow VN$ generated from the same coupled-channel models of $N^*_{c\bar{c}}$.
In the following subsections, we will give formula for calculating these amplitudes.
With the calculated $\Gamma_{N^*_{c\bar{c}},\gamma\,p}$ and
the predicted widths $\Gamma_{N^*_{c\bar{c}},J/\psi\,p}$
and $\Gamma^{(tot)}_{N^*_{c\bar{c}}}$ listed in Table \ref{tab:mwb},
we then can use use Eq.(\ref{eq:R-dy}) to estimate the predicted $\sigma^{(tot)}$ for each model and then select only the cases that the estimated $\sigma^{(tot)}$ are close to the available data to make predictions.

\begin{table}[ht]
      \setlength{\tabcolsep}{0.15cm}
 \begin{center}\caption{The mass ($m$(MeV)), total width ($\Gamma$ (MeV), for $\Gamma_{p\gamma}(kev)$), branch decay width ($\Gamma$) of $J/\psi N$, $\rho N$, $\omega N$, $\bar{D}\Lambda_c$ and $\bar{D}^{*}\Lambda_c$ are listed with various models.The No. 17 and 18 are from the experimental data. In Ref.~\cite{Lin:2017mtz}, the mass is used from experimental data. Others are all calculated from models.}
 \begin{tabular}{cccccccccccc}\hline
No. &    $J^{P}$ & $m$ & $\Gamma$ & $\Gamma_{J/\psi N}$ & $\Gamma_{\rho N}$ & $\Gamma_{\omega N}$   & $\Gamma_{\bar{D} \Lambda_c}$   & $\Gamma_{\bar{D}^* \Lambda_c}$ & $\Gamma_{p\gamma}$ & MC & Ref.       \\
    \hline
1 &     $\frac{1}{2}^-$  & $4262$ & $35.6$  & $10.3$  & $-$  & $-$  & $0.01$ & $-$ & $-$   & $\bar{D}\Sigma_c$  & \cite{Xiao:2013yca}
          \\
2 &                      & $4308$ & $7.1$  & $1.2$  & $-$  & $-$  & $0.02$ & $1.4$ & $-$  & $\bar{D}\Sigma_c$ & \cite{Huang:2018wed}
           \\
3 &                       & $4412$ & $47.3$  & $19.2$  & $3.2$  & $10.4$  & $-$ & $-$ & $-$   & $\bar{D}^*\Sigma_c$& \cite{Wu:2010vk,Wu:2010jy}
     \\
4 &                      & $4410$ & $58.9$  & $52.5$  & $-$  & $-$  & $0.8$ & $0.7$ & $-$  & $\bar{D}^*\Sigma_c$  & \cite{Xiao:2013yca}
        \\
5 &                      & $4460$ & $6.2$  & $3.9$  & $-$  & $-$  & $1.0$ & $0.3$  & $-$   & $\bar{D}^*\Sigma_c$ & \cite{Huang:2018wed}
     \\
6 &                      & $4481$ & $57.8$  & $14.3$  & $-$  & $-$  & $1.02$ & $0.3$ & $-$  & $\bar{D}^*\Sigma^*_c$ & \cite{Xiao:2013yca}
        \\
7 &    $\frac{3}{2}^-$  & $4334$ & $38.8$  & $38.0$  & $-$  & $-$  & $-$ & $0.8$ & $-$   & $\bar{D}\Sigma^*_c$  & \cite{Xiao:2013yca}
       \\
8 &                      & $4375$ & $2.4$  & $1.5$  & $-$  & $-$  & $-$ & $0.9$   & $-$  & $\bar{D}\Sigma^*_c$ & \cite{Huang:2018wed}
     \\
9 &                      & $4380$ & $144.3$  & $3.8$  & $1.4$  & $5.3$  & $1.2$ & $131.3$ & $0.7$ & $\bar{D}\Sigma^*_c$  & \cite{Lin:2017mtz}
      \\
10 &                      & $4380$ & $69.9$  & $16.6$  & $0.15$  & $0.6$  & $17.0$ & $35.3$  & $-$ & $\bar{D}^*\Sigma_c$  & \cite{Lin:2017mtz}
      \\
11 &                      & $4412$ & $47.3$  & $19.2$  & $3.2$  & $10.4$  & $-$  & $-$ & $-$ & $\bar{D}^*\Sigma_c$  & \cite{Wu:2010vk,Wu:2010jy}
       \\
12 &                      & $4417$ & $8.2$  & $4.6$  & $-$  & $-$  & $-$ & $3.1$ & $-$  & $\bar{D}^*\Sigma_c$  & \cite{Xiao:2013yca}
      \\
13 &                      & $4450$ & $139.8$  & $16.3$  & $0.14$  & $0.5$  & $41.4$ & $72.3$ & $-$   & $\bar{D}^*\Sigma_c$  & \cite{Lin:2017mtz}
      \\
14 &                      & $4450$ & $21.7$  & $0.03$  & $-$  & $-$  & $1.4$ & $6.8$ & $-$   & $\bar{D}^*\Sigma_c$ & \cite{Eides:2018lqg}
       \\
15 &                      & $4450$ & $16.2$  & $11$  & $-$  & $-$  & $0.6$ & $4.2$ & $-$   & $\Psi'N$& \cite{Eides:2018lqg}
       \\
16 &                      & $4453$ & $1.8$  & $1.5$  & $-$  & $-$  & $-$ & $0.3$  & $-$   & $\bar{D}\Sigma^*_c$ & \cite{Huang:2018wed}
       \\
17 &                      & $4481$ & $34.7$  & $32.8$  & $-$  & $-$  & $-$ & $1.2$ & $-$  & $\bar{D}^*\Sigma^*_c$ & \cite{Xiao:2013yca}
      \\
18 &     $\frac{5}{2}^+$  & $4450$ & $46.4$  & $4.0$  & $0.3$  & $0.3$  & $18.8$ & $20.5$& $1.13$    & $\bar{D}^*\Sigma_c$  & \cite{Lin:2017mtz}
       \\

 \hline
19 &  $\frac{3}{2}^-$,$\frac{5}{2}^+$
      & $4380^{\pm8}_{\pm29}$ & $205^{\pm18}_{\pm86}$  & $-$  & $-$  & $-$  & $-$ & $-$& $-$   &Exp& \cite{Aaij:2015tga,Aaij:2016ymb}       \\
20 &  & $4450^{\pm2}_{\pm3}$ & $39^{\pm5}_{\pm19}$  & $-$  & $-$  & $-$  & $-$ & $-$   &$-$ & Exp& \cite{Aaij:2015tga,Aaij:2016ymb}       \\
 \hline
 \end{tabular}  \label{tab:mwb}
 \end{center}
 \end{table}

%

\subsubsection{The  $N^*_{c\bar{c}}\to N V$ transition  amplitudes }

Following the formulation of Ref.\cite{Zou:2002yy,Dulat:2011rn},
 the $N^*_{c\bar{c}}(J^P, P)\to N(p_N)+ V(p_V)$ transitions
for spin-parity $J^P=\frac{1}{2}^-, \frac{3}{2}^-$, and
$\frac{5}{2}^+$ can be written as

\begin{eqnarray}
{\cal M}_{N^*(\frac{1}{2}^-)NV}(P;p_V\,p_N)&=&
\bar{u}_N(p_N)\gamma_{5}\tilde{\gamma}_{\mu}u_{N^*}(P)
\epsilon^*_{\nu}(p_V)
\left(
g_{1V}g^{\mu\nu}
-
f_{1V}\left(\frac{3}{2}\frac{\tilde{r}^{\mu}\tilde{r}^{\nu}}{\tilde{r}^2}-\frac{1}{2}\tilde{g}^{\mu\nu}_{N^*}\right)
\right),\label{eq:lagN12NV}\,
\\
{\cal M}_{N^*(\frac{3}{2}^-)NV}(P;p_V\,p_N)&=&
\bar{u}_N(p_N)u_{N^*\,\,\mu}(P)\epsilon^*_{\nu}(p_V)
\left(g_{3V}g^{\mu\nu}
- f_{3V}\left(\frac{3}{2}\frac{\tilde{r}^{\mu}\tilde{r}^{\nu}}{\tilde{r}^2}-\frac{1}{2}\tilde{g}^{\mu\nu}_{N^*}\right)
\right)
\nonumber \\
&&+
h_{3V}\epsilon_{\mu\nu\lambda\delta}\bar{u}_N\gamma_5
(\tilde{\gamma}^{\mu}g_{\alpha}^{\beta}+\tilde{\gamma}_{\alpha}g^{\mu\beta})
u_{N^*\,\,\beta}\epsilon^*_{V\,\,\nu}
\left(\frac{\tilde{r}^{\alpha}
\tilde{r}^{\lambda}}{\tilde{r}^2}
-\frac{1}{3}\tilde{g}^{\alpha\lambda}_{N^*}\right)\hat{P}^{\delta},\label{eq:lagN32NV},\,
\\
{\cal M}_{N^*(\frac{5}{2}^+)NV}(P;p_V\,p_N)&=&
\bar{u}_N(p_N) u_{N^*\,\,\mu\nu}(P)\epsilon^*_{\alpha}(p_V)\nonumber \\
&&\times
\left(\frac{g_{5V}}{m_N}g^{\alpha\mu}\tilde{r}^{\nu}
- \frac{f_{5V}}{m_N}\left(\frac{5}{3}\frac{\tilde{r}^{\mu}\tilde{r}^{\nu}\tilde{r}^{\alpha}}{\tilde{r}^2}
-\frac{1}{3}\left(
\tilde{g}^{\mu\nu}_{N^*}\tilde{r}^{\alpha}
+\tilde{g}^{\nu\alpha}_{N^*}\tilde{r}^{\mu}
+\tilde{g}^{\alpha\mu}_{N^*}\tilde{r}^{\nu}\right)\right)
\right)
\nonumber \\
&&
+
\frac{h_{5V}}{m_N}\epsilon_{\mu\nu\lambda\delta}\bar{u}_N\gamma_5
\left(\tilde{\gamma}^{\mu}g_{\xi\alpha}g_{\sigma\beta}
+\tilde{\gamma}_{\xi}g_{\sigma\beta}g_{\mu\beta}
+\tilde{\gamma}_{\sigma}g_{\mu\beta}g_{\xi\beta}
\right)
u_{N^*}^{\alpha\beta}\epsilon_{V}^{*\,\,\mu}
\nonumber\\&&
\times
\left(\frac{\tilde{r}^{\xi}\tilde{r}^{\lambda}\tilde{r}^{\sigma}}{\tilde{r}^2}
-\frac{1}{3}\left(
\tilde{g}^{\xi\sigma}_{N^*}\tilde{r}^{\lambda}
+\tilde{g}^{\sigma\lambda}_{N^*}\tilde{r}^{\xi}
+\tilde{g}^{\lambda\xi}_{N^*}\tilde{r}^{\sigma}\right)\right)\hat{P}^{\delta}
,\label{eq:lagN52NV},\,
\end{eqnarray}
where
\begin{eqnarray}
r^{\mu}&=&p^{\mu}_N - p^{mu}_V,\,\,\,\,\,\,\,\,\,
\tilde{g}^{\mu\nu}_{N^*}=g^{\mu\nu}-\frac{P^\mu P^\nu}{W^2},\,\label{eq:tilde-1}\\
\tilde{r}^\nu&=&r_{\mu}\tilde{g}^{\mu\nu}_{N^*},\,\,\,\,\,\,\,\,\,\,\,\,\,\,\,\,\,\,
\tilde{\gamma}^\nu=\gamma_{\mu}\tilde{g}^{\mu\nu}_{N^*},\, \label{eq:tilde-2}\\
\hat{P}_{\mu}&=&P_{\mu}/W
\label{eq:lagN12NV-f}
\end{eqnarray}

The terms with coupling constants $f_{2J\,V}$ and $h_{2J\,V}$ term are the contributions from
 higher partial waves. For simplicity, we neglect these terms
and set $f_{2J\,V}=h_{2J\,V}=0$. We thus can use the partial decay width $\Gamma_{N^*_{c\bar{c}}, NV }$
listed in Table \ref{tab:mwb} to determine the
parameter $g_{2J\,V}$ by the following formula:
\begin{eqnarray}
\Gamma_{N^*_{c\bar{c}}NV}
=\frac{|p|}{8\pi m^2_{N^*_{c\bar{c}}}}\left|{\cal M}_{N^*_{c\bar{c}}NV}\right|^2,
\label{eq:ns-vn}
\end{eqnarray}
where $|p|$ is on-shell momentum of final state vector in the rest frame of $N^*_{c\bar{c}}$.

 The determined $g_{2J\,V}$ for $V=J/\psi$ a
re listed in the 5th column of Table \ref{tab:coup-g}.

 \begin{table}[ht]
      \setlength{\tabcolsep}{0.15cm}
 \begin{center}\caption{coupling constants $g_V$ and $\tilde{g}_V$ determined from
fitting the partial decay widths listed in Table \ref{tab:mwb}.
$\Gamma_{\gamma p}$ are partial decay widths calculated from $\tilde{g}_V$ within VMD,
as explained in the text.
$\sigma^{(tot)}$ is  the total cross section
of $\gamma\,p\rightarrow J/\psi\, p$ calculated from using Eq.(20) by choosing
$\Lambda=0.55$ GeV for the off-shell form factor $F_V(q^2)$.}
 \begin{tabular}{ccccccccccccc}\hline
No. &    $J^{P}$ & $m$  & $\Gamma_{tot}$ & ${g}_{2J\,V}$
&  $\tilde{g}_{2J\,V}$ & $\tilde{g}_{2J\,V}$
 & $\tilde{g}_{2J\,V}$    & $\Gamma_{p\gamma}$ (kev) &  $\sigma^{(tot)}$(nb)    \\
  &     &   &   & $J/\psi p$ &  $J/\psi p$ & $\rho p$
 & $\omega p$    &  &      \\
    \hline
1 &  $\frac{1}{2}^-$  & $4262$ & $35.6$  & $0.39$  & $0.32$  & $-$      & $-$
                      & $3.9\times10^{-5}$         & $1.9\times10^{-4}$   \\
2 &                   & $4308$ & $7.1$     & $0.13$  & $0.11$  & $-$      & $-$
                      & $4.5\times10^{-6}$         & $5.9\times10^{-9}$   \\
3 &                   & $4412$ & $47.3$  & $0.46$  & $0.38$  & $0.078$  & $0.14$
                      & $1.14$           & $5.4$   \\
4 &                   & $4410$ & $58.9$  & $0.75$  & $0.62$  & $-$      & $-$
                      & $1.5\times10^{-4}$         & $1.3\times10^{-3}$   \\
5 &                   & $4460$ & $6.2$     & $0.20$  & $0.16$  & $-$      & $-$
                      & $1.1\times10^{-5}$         & $6.2\times10^{-4}$   \\
6 &                   & $4481$ & $57.8$  & $0.37$  & $0.31$  & $-$      & $-$
                      & $3.8\times10^{-5}$         & $8.8\times10^{-5}$   \\
7 &  $\frac{3}{2}^-$  & $4334$ & $38.8$  & $1.19$  & $0.98$  & $-$      & $-$
                      & $1.3\times10^{-4}$         & $3.7\times10^{-3}$   \\
8 &                   & $4375$ & $2.4$     & $0.23$  & $0.19$  & $-$      & $-$
                      & $4.6\times10^{-6}$         & $1.4\times10^{-3}$   \\
9 &                   & $4380$ & $144.3$ & $0.36$  & $0.30$  & $0.090$  & $0.17$
                      & $0.53$                     & $0.11$   \\
10 &                  & $4380$ & $69.9$  & $0.75$  & $0.62$  & $0.039$  & $0.059$
                      & $0.060$                    & $0.23$   \\
11 &                  & $4412$ & $47.3$  & $0.79$  & $0.65$  & $0.14$   & $0.24$
                      & $1.1$                      & $10.8$   \\
12 &                  & $4417$ & $8.2$   & $0.39$  & $0.32$  & $-$      & $-$
                      & $1.4\times10^{-5}$         & $1.0\times10^{-3}$   \\
13 &                  & $4450$ & $139.8$ & $0.71$  & $0.58$  & $0.028$  & $0.053$
                      & $0.054$                    & $0.048$   \\
14 &                  & $4450$ & $21.7$  & $0.030$ & $0.025$ & $-$      & $-$
                      & $8.4\times10^{-8}$         & $5.8\times10^{-9}$   \\
15 &                  & $4450$ & $16.2$  & $0.58$  & $0.48$  & $-$      & $-$
                      & $3.1\times10^{-5}$         & $1.4\times10^{-3}$   \\
16 &                  & $4453$ & $1.8$     & $0.21$  & $0.18$  & $-$      & $-$
                      & $4.2\times10^{-6}$         & $2.2\times10^{-3}$   \\
17 &                  & $4481$ & $34.7$  & $0.98$  & $0.81$  & $-$      & $-$
                      & $8.8\times10^{-5}$         & $2.6\times10^{-3}$   \\
18 & $\frac{5}{2}^+$  & $4450$ & $46.4$  & $0.35$  & $0.27$  & $0.016$  & $0.016$
                      & $8.3\times10^{-2}$         & $0.25$   \\
    \hline
 \end{tabular}  \label{tab:coup-g}
 \end{center}
 \end{table}

\subsubsection{The  $N^*_{c\bar{c}}\to N\gamma$ transition amplitudes }

As illustrated in Fig.\ref{fg:ncc}, we assume that the  $N^*_{c\bar{c}}\to N\gamma$ transition amplitudes can be calculated by the $\gamma \rightarrow V$ transition defined by
the VMD Lagrangian  Eq.(\ref{eq:l-vmd}), the propagator of $V$, and the $VN \rightarrow N^*_{c\bar{c}}$ amplitudes defined in Eqs.(\ref{eq:lagN12NV})-(\ref{eq:lagN12NV-f}).
Since we can determine the parameters by using only one value of $\Gamma_{N^*_{c\bar{c},\gamma N}}$ predicted by a $N^*_{c\bar{c}}$ model, we need to make simplification.
Here we also need to make sure that the simplified amplitudes are gauge invariant.
We find that this can be accomplished by setting $h_{2J\,V}=0$
like what we have chosen in determining $\Gamma_{N^*_{c\bar{c}}}\rightarrow NV$,
but we  need to  keep the $f_{2J\,V}$ term and set $g_{2J\,V}=f_{2J\,V}=\tilde{g}_{2J\,V}$.
For example, the amplitude of  $N^*(\frac{1}{2}^-) \to N\rho \to N\gamma$ with the simplification $h_{2J\,V}=0$ is:
\begin{eqnarray}
{\cal M}_{N^*(\frac{1}{2}^-)\to N\rho\to N\gamma}(P;q=p_{\rho}\,p_N)
= \frac{ie}{f_{\rho}}\frac{-m^2_{\rho}}
{q^2-m^2_{\rho}+i\Gamma_{\rho}m_{\rho}}\mathcal{M}^{\nu}(P;p_\rho\,p_N)
\epsilon^{*\,\,\nu}_{\gamma}(q) \nonumber \\
\end{eqnarray}
with
\begin{eqnarray}
\mathcal{M}^{\nu}(P;q=p_{\rho}\,p_N)=
 \overline{u}_{N^*}(P)\gamma_{5}\tilde{\gamma}_{\mu}u_N(p_N)
\left(
g_{1\rho}g^{\mu\nu'}
-
f_{1\rho}\left(\frac{3}{2}\frac{\tilde{r}^{\mu}\tilde{r}^{\nu}}{\tilde{r}^2}-\frac{1}{2}\tilde{g}^{\mu\nu'}_{N^*}\right)\right)\tilde{g}_{\rho\,\,\nu'\nu}(q)
\nonumber \\
&&
\end{eqnarray}
Obviously this amplitude will be gauge invariant if $\mathcal{M}^{\nu}q_{\nu}=0$.
However it is straightforward to show that  $\mathcal{M}^{\nu}q_{\nu}\sim (g_{1\rho}-f_{1\rho})\neq 0$.
Therefore a  simple way to have a gauge invariant amplitude is to set $g_{2J\,V}=f_{2J\,V}=\tilde{g}_{2J\,V}$.
This is part of phenomenology and need to be improved in future.
For our present limited and exploratory purpose, this simplification is sufficient.

By using Eqs.(\ref{eq:lagN12NV})-(\ref{eq:lagN52NV}) and setting $g_{2J\,V}=f_{2J\,V}=\tilde{g}_{2J\,V}$
and $h_{2J\,V}=0$, we can then use Eq.(\ref{eq:ns-vn}) to
determine $\tilde{g}_{2J\,V}$ by using the partial decay withs listed in Table \ref{tab:mwb}.
The resulting $\tilde{g}_{2J\,V}$ are listed in the 6th-8th columns of Table \ref{tab:coup-g}.
Including the off-shell form factor $F_V(q^2)$ according to Eq.(\ref{eq:off-shell}), we then get the following
expressions for the $N^*_{c\bar{c}}\rightarrow VN \rightarrow \gamma N$ transition amplitudes:
\begin{eqnarray}
{\cal M}_{N^*(\frac{1}{2}^-)\to NV\to N\gamma}(P;q\,p_N)&=&
\frac{ie}{f_{V}}
\frac{-m^2_{V}\tilde{g}_{1V}}
{-m^2_{V}+i\Gamma_{V}m_{V}}
\bar{u}_{N}(p_N)\gamma_{5}\tilde{\gamma}_{\mu}u_{N^*}(P)\nonumber \\
&&\times
[\epsilon^*_{\gamma\,\,\nu}
\left(
g^{\mu\nu}
-
\frac{3}{2}\frac{\tilde{r}^{\mu}\tilde{r}^{\nu}}{\tilde{r}^2}
+\frac{1}{2}\tilde{g}^{\mu\nu}_{N^*}\right)F_V(q^2)]
\label{eq:lagN12Ng},\,
\\
{\cal M}_{N^*(\frac{3}{2}^-)\to NV\to N\gamma}(P;q\,p_N)&=&
\frac{ie}{f_{V}}
\frac{-m^2_{V}\tilde{g}_{3V}}
{-m^2_{V}+i\Gamma_{V}m_{V}}\bar{u}_{N}(p_N)u_{N^*\,\,\mu}(P)\nonumber \\
&&\times[\epsilon^*_{\gamma\,\,\nu}
\left(g^{\mu\nu}
-\frac{3}{2}\frac{\tilde{r}^{\mu}\tilde{r}^{\nu}}{\tilde{r}^2}+\frac{1}{2}\tilde{g}^{\mu\nu}_{N^*}\right)
F_V(q^2)]
\label{eq:lagN32Ng},\,
\\
{\cal M}_{N^*(\frac{5}{2}^+)\to NV\to N\gamma}(P;q\,p_N)&=&
\frac{ie}{f_{V}}
\frac{-m^2_{V}\tilde{g}_{5V}/m_N}
{-m^2_{V}+i\Gamma_{V}m_{V}}
\bar{u}_{N}(p_N)u_{N^*\,\,\mu\nu}(P)\epsilon^*_{\gamma\,\,\alpha}
F_V(q^2)
\nonumber\\&&\times
\left(g^{\alpha\mu}\tilde{r}^{\nu}
-\frac{5}{3}\frac{\tilde{r}^{\mu}\tilde{r}^{\nu}\tilde{r}^{\alpha}}{\tilde{r}^2}
+\frac{1}{3}\left(
\tilde{g}^{\mu\nu}_{N^*}\tilde{r}^{\alpha}
+\tilde{g}^{\nu\alpha}_{N^*}\tilde{r}^{\mu}
+\tilde{g}^{\alpha\mu}_{N^*}\tilde{r}^{\nu}\right)
\right)\label{eq:lagN52Ng}.
\end{eqnarray}
where $p_V=q$ is used to evaluate $\tilde{r}^{\nu}$ and $\tilde{g}^{\alpha\mu}_{N^*}$
according to Eqs.(\ref{eq:tilde-1})-(\ref{eq:tilde-2}).
For the off-shell form factors, we assume
\begin{eqnarray}
F_V(q) &=& \frac{\Lambda^4_V}{\Lambda^4_V+(q^2-m^2_V)^2}.
\label{eq:off-ff}
\end{eqnarray}

With the determined $\tilde{g}_V$ listed in Table \ref{tab:coup-g} and a given
 choice of the
cut off $\Lambda_V$, we can use the $N^* \to NV \to N\gamma$ amplitudes given in
Eqs.(\ref{eq:lagN12Ng})-(\ref{eq:lagN52Ng})
to calculate the decay width of ${N^* \to N\gamma}$ within VMD  :
\begin{eqnarray}
\Gamma_{N^* \to N\gamma} &=& \frac{1}{8\pi}\frac{k}{m^2_{N^*}}\overline{\left|\sum_{V}M_{N^* \to NV \to N\gamma}\right|^2}.
\end{eqnarray}
The cut-off $\Lambda$ is a parameter of the model.
In Table \ref{tab:coup-g} we list the calculated $\Gamma_{N^* \to N\gamma}$ for each model by setting $\Lambda=0.55$ GeV(The dependence on the value of $\Lambda$ will be discussed in the next section).
By using the partial decay widths listed in
Tables \ref{tab:mwb} and \ref{tab:coup-g} we can use Eq.(\ref{eq:R-dy}) to estimate
the total cross section $\sigma^{(tot)}$  of $\gamma\,p\rightarrow J/\psi\,p$ at the resonance positions,
as also given in the 9th column of  Table \ref{tab:coup-g}.


\subsubsection{The amplitude of $\gamma p \to N^*_{c\bar{c}} \to J/\psi p $ }

The amplitude $\gamma p \to N^*_{c\bar{c}} \to J/\psi p $ is shown in Fig.\ref{fg:ncc}.
By using the definition of vertexes of $N^* \to NV$ as shown Eq.(\ref{eq:lagN12NV}-\ref{eq:lagN52NV}) and $N^*\to NV \to N\gamma$ as shown Eq.(\ref{eq:lagN12Ng}-\ref{eq:lagN52Ng}), we can write the amplitude $\mathcal{M}_{N^*}^{\mu\nu}(q,p,q',p')$ which defined in Eq.(\ref{eq:ampjpsip}):
\begin{eqnarray}
\mathcal{M}_{N^*(\frac{1}{2}^-)}^{\mu\nu}(q,p,q',p') &=&
\sum_{V=J/\psi,\rho,\omega}g_{1J/\psi}\gamma_{5}\tilde{\gamma}_{\alpha}
\tilde{g}^{\alpha\mu}(q)
\frac{\slashed{q}+\slashed{p}+m_{N^*_{c\bar{c}}}}
{W^2-m^2_{N^*_{c\bar{c}}}+i\Gamma_{N^*_{c\bar{c}}}m_{N^*_{c\bar{c}}}}
F_V(0)
\nonumber\\&&\times
\frac{ie}{f_{V}}
\frac{-m^2_{V}\tilde{g}_{1V}}
{-m^2_{V}+i\Gamma_{V}m_{V}}
\gamma_{5}\tilde{\gamma}_{\beta}
\left(
g^{\beta\nu}
-
\frac{3}{2}\frac{\tilde{r}^{\beta}\tilde{r}^{\nu}}{\tilde{r}^2}
+\frac{1}{2}\tilde{g}^{\beta\nu}_{N^*}\right),
\label{eq:famp-1}
\\
\mathcal{M}_{N^*(\frac{3}{2}^-)}^{\mu\nu}(q,p,q',p') &=& \sum_{V=J/\psi,\rho,\omega}g_{3J/\psi}g^{\mu\alpha}
\frac{(\slashed{q}+\slashed{p}+m_{N^*_{c\bar{c}}})P_{\alpha\beta}^{\frac{3}{2}}(p+q)}
{W^2-m^2_{N^*_{c\bar{c}}}+i\Gamma_{N^*_{c\bar{c}}}m_{N^*_{c\bar{c}}}}F_V(0)
\nonumber\\&&\times
\frac{ie}{f_{V}}
\frac{-m^2_{V}\tilde{g}_{3V}}
{-m^2_{V}+i\Gamma_{V}m_{V}}
\left(
g^{\beta\nu}
-\frac{3}{2}\frac{\tilde{r}^{\beta}\tilde{r}^{\nu}}{\tilde{r}^2}
+\frac{1}{2}\tilde{g}^{\beta\nu}_{N^*}\right),
\\
\mathcal{M}_{N^*(\frac{5}{2}^+)}^{\mu\nu}(q,p,q',p') &=&\sum_{V=J/\psi,\rho,\omega}
\frac{g_{5J/\psi}}{m_N}g^{\mu\alpha}\tilde{r}^{\beta}
\frac{(\slashed{q}+\slashed{p}+m_{N^*_{c\bar{c}}})
P_{\alpha\beta\,\alpha'\beta'}^{\frac{5}{2}}(p+q)}
{W^2-m^2_{N^*_{c\bar{c}}}+i\Gamma_{N^*_{c\bar{c}}}m_{N^*_{c\bar{c}}}}
 F_V(0)
\nonumber\\&&\times
\frac{ie}{f_{V}}
\frac{-m^2_{V}\tilde{g}_{5V}/m_N}
{-m^2_{V}+i\Gamma_{V}m_{V}}
\nonumber\\&&\times
\left(g^{\nu\alpha'}\tilde{r}^{\beta'}
-\frac{5}{3}\frac{\tilde{r}^{\nu}\tilde{r}^{\alpha'}\tilde{r}^{\beta'}}{\tilde{r}^2}
+\frac{1}{3}\left(
\tilde{g}^{\nu\alpha'}_{N^*}\tilde{r}^{\beta'}
+\tilde{g}^{\nu\beta'}_{N^*}\tilde{r}^{\alpha'}
+\tilde{g}^{\alpha'\beta'}_{N^*}\tilde{r}^{\nu}\right)
\right),
\label{eq:famp-3}
\end{eqnarray}
where $P_{\alpha\beta}^{\frac{3}{2}}(p)$ and $P_{\alpha\beta\,\alpha'\beta'}^{\frac{5}{2}}(p)$, are the Lorentz structure functions of propagators of $3/2$ and $5/2$ particles, respectively. Their formulas are \cite{Dulat:2011rn}:
\begin{eqnarray}
P_{\alpha\beta}^{\frac{3}{2}}(p)&=&
-g_{\alpha\beta}+\frac{1}{3}\gamma_{\mu}\gamma_{\nu}+\frac{2}{3}\frac{p_{\mu}p_{\nu}}{m^2_{N^*}}+\frac{1}{3m_{N^*}}(\gamma_{\mu}p_{\nu}-\gamma_{\nu}p_{\mu})
\\
P^{\frac{5}{2}\,\,\alpha\beta\,\alpha'\beta'}(p)&=&
\frac{1}{2}(\tilde{g}^{\alpha\alpha'}_{N^*}\tilde{g}^{\beta\beta'}_{N^*}
+\tilde{g}^{\alpha\beta'}_{N^*}\tilde{g}^{\beta\alpha'}_{N^*})
-\frac{1}{5}\tilde{g}^{\alpha\beta}_{N^*}\tilde{g}^{\alpha'\beta'}_{N^*}
\nonumber\\&&
-\frac{1}{10}\left(
\tilde{\gamma}^{\alpha}\tilde{\gamma}^{\alpha'}\tilde{g}^{\beta\beta'}_{N^*}
+\tilde{\gamma}^{\alpha}\tilde{\gamma}^{\beta'}\tilde{g}^{\beta\alpha'}_{N^*}
+\tilde{\gamma}^{\beta}\tilde{\gamma}^{\alpha'}\tilde{g}^{\alpha\beta'}_{N^*}
+\tilde{\gamma}^{\beta}\tilde{\gamma}^{\beta'}\tilde{g}^{\alpha\alpha'}_{N^*}
\right)
\label{eq:famp-f}
\end{eqnarray}

\begin{figure}[htbp] \vspace{-0.cm}
\begin{center}
\includegraphics[width=1.0\columnwidth,angle=0]{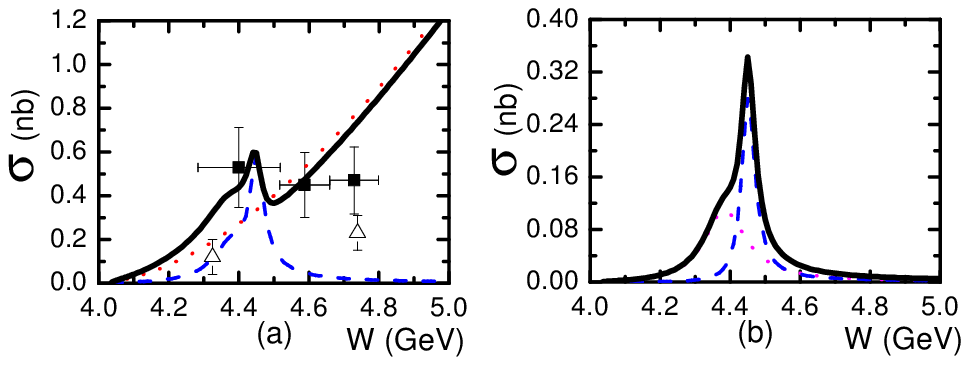}
\caption{The total cross sections of $\gamma p \to J/\psi p$ diagram with the invariant mass of $\gamma p$.
(a) The red dotted and blue dashed curves are the contribution from the
 Pomeron-exchange  and $N^*_{c\bar{c}}$ with $J^p$= $3/2^-$ and $5/2^+$ with $\Lambda=550$ MeV, respectively.
The black solid line is for the coherent summation of all above three contributions.
(b) The red dotted and blue dashed curves are the contribution of $N^*_{c\bar{c}}$ with $J^p$= $3/2^-$ and $5/2^+$ with $\Lambda=550$ MeV, respectively.
The solid black line is the coherent summation of two $N^*{c\bar{c}}$ and it is the same as blue dashed cure in Fig.(a) here.
The experimental data is from Refs.~\cite{Gittelman:1975ix,Anderson:1976sd}.
}
\label{fg:totaljpsip}
\end{center}
\end{figure}

\section{Predictions for $\gamma p \rightarrow J/\psi p$ }

In this section, we will first use the available total cross section data to fix the cutoff parameter $\Lambda$ of the off-shell form factor Eq.(\ref{eq:off-ff}) of the $\gamma p \rightarrow N^*_{c\bar{c}}$ amplitude.
We then make predictions for using differential cross sections for identifying
the $ N^*_{c\bar{c}}$ from the future experimental data.

\subsection{Total cross section}

From Fig.\ref{fg:backg}, we see that the available data of the $\gamma\, p \rightarrow J/\psi\, p$ in the near threshold region are below about  0.8 $nb$ and have some structure which may be due to the experimental uncertainties, but may be due to the $N^*_{c\bar{c}}$ excitations.
In this section we will make predictions for investigating the extent to which these available data can accommodate the the $N^*_{c\bar{c}}$ excitations predicted by the models listed in Table I and II.
In particular, we are interested in the predictions of Ref.\cite{Lin:2017mtz} since this is the only model which predicts the partial decay width to $\gamma p$ channel for the $\frac{3}{2}^-$(4380) and $\frac{5}{2}^+$(4450) states.
The tri-angular mechanism they used for the $\gamma N \rightarrow N^*_{c\bar{c}}(4380)$ is similar as our model based on VMD, but for $N^*_{c\bar{c}}(4450)$ they are different with one magnitude order.

Our first step is to determine the cutoff parameter $\Lambda$ of the off-shell form factor Eq.(\ref{eq:off-ff}).
To compare with the results of Ref.\cite{Lin:2017mtz},
we perform  calculations including
 only $\frac{3}{2}^-$(4380) and $\frac{5}{2}^+$(4450)
using the parameters  (No. 9 and 18 of Ref.\cite{Lin:2017mtz}) listed in Table \ref{tab:coup-g}.
We find that the calculated total cross sections
can be close to the available data shown in Fig.\ref{fg:backg}
 if we choose the cutoff in the range of $500$ MeV $\leq \Lambda\leq 650$ MeV.
In Fig.\ref{fg:totaljpsip}(a), we see that the choice $\Lambda=550$ MeV  gives
 results within the uncertainties of the available data. The structure of the solid
curve at $W\sim 4.35$ GeV is due to the interference between the
Pomeron-exchange amplitude (dotted curve) and the resonant amplitude (long dashed curve).
Furthermore, we also see that the resonant amplitude is dominated by the
$\frac{5}{2}^+$(4450), as shown in Fig.\ref{fg:totaljpsip}(b).

With the same cutoff $\Lambda=550$ MeV , we then calculate $\Gamma_{N^*_{c\bar{c}}\rightarrow \gamma N}$
for all states, as
listed in Table \ref{tab:coup-g}.  With the widths given in Table \ref{tab:coup-g}, we then estimate the total
cross sections of $\gamma p \rightarrow N^*_{c\bar{c}}\rightarrow J/\psi p$ by using
Eq.(\ref{eq:R-dy}) for all models.
We can see in the last column of Table \ref{tab:coup-g} that except
the $\frac{3}{2}^-$(4380) and $\frac{5}{2}^+$(4450) of Ref.\cite{Lin:2017mtz},
all of the estimated total cross sections are either too large or too small
compared with the value $\sim 0.5 $ nb of the available data shown in Fig.\ref{fg:totaljpsip}.

\begin{table}[ht]
      \setlength{\tabcolsep}{0.15cm}
\begin{center}\caption{
fitting the partial decay widths of the states included in our predictions.
$\Gamma_{\gamma p}$ are partial decay widths calculated from $\tilde{g}_V$ within VMD,
as explained in the text.
$\sigma^{(tot)}$ is  the total cross section
of $\gamma\,p\rightarrow J/\psi\, p$ calculated from using Eq.(20) by choosing
$\Lambda=0.55$ GeV for the off-shell form factor $F_V(q^2)$.}
 \begin{tabular}{ccccccccccc}\hline
No. &    $J^{P}$ & $m$  & $\Gamma_{tot}$ & $\Gamma_{J/\psi p}$
 & $\Gamma_{p\gamma}$ (kev) &  $\sigma^{(tot)}$(nb)& Ref.    \\
    \hline
9 &  $\frac{3}{2}^-$  & $4380$ & $144.3$ & $3.8$  & $0.53$                     & $0.11$ & This work \\
  &                   & $4380$       & $144,3$      &   $3.8$       & $0.70$                     & $0.15$
& \cite{Lin:2017mtz} \\
18 & $\frac{5}{2}^+$  & $4450$ & $46.4$  & $4.0$  & $0.083$         & $0.25$ & This work  \\
   &                  & $4450$       & $46.4$        &  $4.0$      & $1.13$           & $3.4$ &\cite{Lin:2017mtz} \\
    \hline
 \end{tabular}  \label{tab:jpsi-cal}
 \end{center}
 \end{table}

\begin{figure}[htbp] \vspace{-0.cm}
\begin{center}
\includegraphics[width=1.0\columnwidth,angle=0]{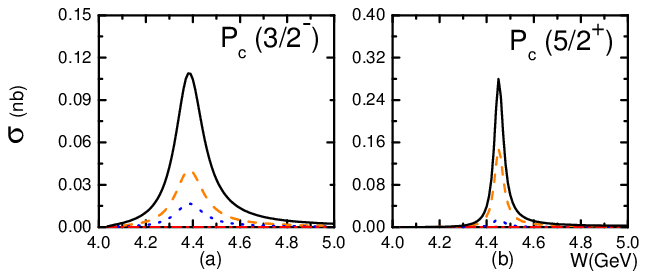}
\caption{
The cross sections of $\gamma p \to N^*_{c\bar{c}} \to J/\psi p $ diagram for different $J^P$ of $N^*_{c\bar{c}}$ with the invariant mass of $\gamma p$.
The orange dashed, blue dotted and red dashed-dotted lines are the contribution purely from the VMD by $\rho$, $\omega$ and $J/\psi$ coupled with $\gamma$, respectively.
The black solid lines are for the coherent summation of $\rho$, $\omega$ and $J/\psi$ contributions.
}
\label{fg:res1GeV}
\end{center}
\end{figure}

In Table \ref{tab:jpsi-cal}, we compare our results of $\Gamma_{\gamma p}$ and
$\sigma^{(tot)}$ of $\gamma p \rightarrow N^*_{c\bar{c}}\rightarrow J/\psi p$ with those of
Ref.\cite{Lin:2017mtz}. Here we see that our result for the $\frac{5}{2}^+$(4450) is much smaller than theirs.
The differences between this work and Ref.~\cite{Lin:2017mtz} are from using rather different
mechanisms  to evaluate $\gamma N\rightarrow N^*_{c\bar{c}}$.
It is therefore useful to examine how our predictions depend on the parameters of our  model based on VMD.
We first examine the the contribution from each of the intermediate vector  mesons, illustrated in Fig.\ref{fg:ncc},  to the calculated total cross sections of $\gamma\, p \rightarrow N^*_{c\bar{c}}\rightarrow J/\psi\,p$.
Our results from including the $J=\frac{3}{2}^-$(4380) and $\frac{5}{2}^+$(4450) in the calculation are shown in  Fig.\ref{fg:res1GeV}.
Clearly the intermediate $\rho$ gives the largest contribution, and $J/\psi$ is negligible.
This can be understood from the employed off-shell form factor Eq.(\ref{eq:off-ff}) which depends on the mass of the intermediate vector meson.
This is also the reason why
 the cross sections predicted by the  models without $\rho p$ channel
listed in Table.\ref{tab:coup-g} are extremely small.

%
%

%
%

%
%
%

%

\subsection{differential cross sections}
\begin{figure}[htbp] \vspace{-0.cm}
\begin{center}
\includegraphics[width=0.6\columnwidth,angle=0]{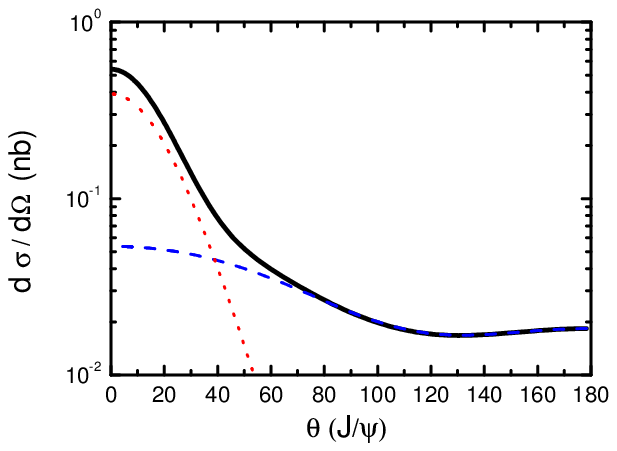}
\caption{The differential cross sections of $\gamma p \to J/\psi p$ diagram with the angular of outgoing $J/\psi$ at invariant mass of $\gamma p$ $W=4.45$ GeV.
The red dotted, blue dashed lines are the contribution purely from the Pomeron and $N^*_{c\bar{c}}$ with $\Lambda=0.55$ GeV, respectively.
The black solid line is for the coherent summation of all contributions.
}
\label{fg:cut}
\end{center}
\end{figure}

\begin{figure}[htbp] \vspace{-0.cm}
\begin{center}
\includegraphics[width=1.0\columnwidth,angle=0]{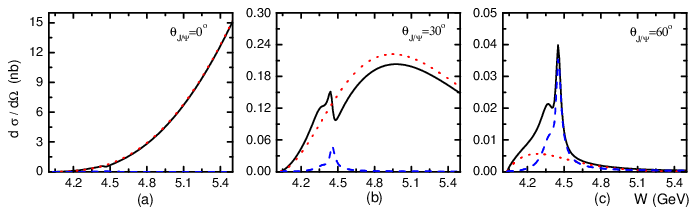}
\caption{The differential cross sections of $\gamma p \to J/\psi p$ diagram with invariant mass of $\gamma p$ at three fixed angular of outgoing $J/\psi$ .
The red dotted, blue dashed lines are the contribution purely from the Pomeron-exchange and $N^*_{c\bar{c}}$ with $\Lambda=0.55$ GeV, respectively.
The black solid line is for the coherent summation of all contributions.
}
\label{fg:cut-ang}
\end{center}
\end{figure}

In Fig.\ref{fg:totaljpsip}, we see that the feature  of $N^*_{c\bar{c}}$ excitation in the total cross section is not so pronounced because it interfere with the background form Pomeron-exchange amplitude which is  very large in all energy region.
To extract the peak of $N^*_{c\bar{c}}$, we need to find other observables which
are not dominated by  the Pomeron exchange.
It is noticed that the Pomeron exchange is strongly suppressed with large $t$ in Eq.(\ref{eq:regge-g}).
In other word, the Pomeron-exchange mainly contribute to the cross sections at forward angles.
This is illustrated in Fig.\ref{fg:cut}.
It is then clear that the  resonance peaks will be easier to observe at large angles.
This is illustrated in Fig.\ref{fg:cut-ang}.
At $60^o$, the shoulder due to $N^*_{c\bar{c}}$(4380) shows up more clearly.
However, the magnitudes of the differential cross sections decrease rather rapidly with angles.
Thus the measurement around $30^o$ may be optimal in examining the existence $N^*_{c\bar{c}}$.

\section{Prediction on  $\gamma p \to \bar{D}^0 \Lambda^+_c,  \bar{D}^{*\,0} \Lambda^+_c $}

It is important to note that Pomeron-exchange amplitude is still dominant in determining the $J/\psi$ production in the considered low energy region.
Therefore it is interesting to test the VMD model of $\gamma N\rightarrow N^*_{c\bar{c}}$ by
other reactions which do not have Pomeron-exchange mechanism and in the low energy region accessible to
experiments measuring $J/\psi$ production at JLab.
With the $N^*_{c\bar{c}}$  models No. 6, 9, and 18 selected from
Table \ref{tab:mwb} and  listed in Table \ref{tab:mwb2}, the reaction
$\gamma\, p\rightarrow \bar{D}^{*\,0} \Lambda^+_c\,, \bar{D}^{0} \Lambda^+_c\,$
can be used for this purpose.
In addition to calculating the
$\gamma\,p\rightarrow N^*_{c\bar{c}}\rightarrow \bar{D}^{*\,0} \Lambda^+_c\,, \bar{D}^{0} \Lambda^+_c\,$
amplitude,
 we also need to consider the meson-exchange mechanisms
due to $\bar{D}^{*0}\rightarrow \bar{D}^0 \gamma $ process.
We thus need to calculate  the amplitudes of the  two  mechanisms shown in Fig. \ref{fig:ndlc}.

\subsection{meson-exchange amplitude}

The meson-exchange amplitudes shown in Fig. \ref{fig:ndlc} (b) can be calculated by using
 $D^* D\gamma$, $p D^{*0}\Lambda_c^+$ and $p D^{0}\Lambda_c^+$ vertices defined as follows:
\begin{eqnarray}
{\cal M}_{D^*D\gamma}&=&\frac{g_{D^{*0}D^{0}\gamma}}{\sqrt{m_{D^*}m_D}}
\epsilon^{\mu\nu\alpha\beta}P_{\gamma\,\mu}\epsilon_{\gamma\,\nu}
P_{D^*\,\alpha}\epsilon_{D^*\,\nu}
\label{eq:dd1}\,,\,
\\
{\cal M}_{ND\Lambda^+_c}&=&
g_{ND\Lambda_c^+}\,\bar{u}_{\Lambda^+_c}\gamma_5 u_{N}
\label{eq:dd2}\,,
\\
{\cal M}_{ND^*\Lambda^+_c}&=&
g_{ND^*\Lambda_c^+}\,\bar{u}_{\Lambda^+_c}\gamma_{\mu} \epsilon^\mu_{D^*} u_{N}
\label{eq:dd3}\,,
\end{eqnarray}
where coupling $g_{D^{*0}D^{0}\gamma} = 1.07$ is calculated from partial decay width of
$D^{*0}\to D^{0}\gamma$ which is estimated from the measured ratio of widths
$\Gamma_{D^{*0}\to D^{0}\gamma}/\Gamma_{D^{*0}\to D^{0}\pi^0}$ with $\Gamma_{D^{*0}\to D^{0}\pi^0}$
obtained from the data of  $\Gamma_{D^{*+}\rightarrow D^{+}\pi^0}$ by using isospin  .
 By using SU(4) symmetry~\cite{Shen:2019evi}, the coupling constants in Eqs.(\ref{eq:dd2})-(\ref{eq:dd3})
can be determined :
$g_{ND\Lambda_c^+}=-\frac{3\sqrt{3}}{5}g_{BBP}$
and $g_{ND^*\Lambda_c^+}=-\sqrt{3}g_{BBV}$, where $g_{BBP}=0.989$ and $g_{BBV}=3.25$.
Then the amplitude $j^{\nu}_{D^*\,\,m'_s,\,m_s}(q,p,q',p')$, defined in Eq.(\ref{eq:tot-amp}),
for $\gamma p \to \bar{D}^0 \Lambda^+_c$ due to $\bar{D}^{*0}$-exchange can be written as
\begin{eqnarray}
j^{\nu}_{D^*\,\,m'_s,\,m_s}(q,p,q',p')
&=&\frac{g_{D^*D\gamma}g_{ND^*\Lambda_c^+}}{\sqrt{m_{D^*}m_D}}
\epsilon^{\mu\nu\alpha\beta}
\frac{q_{\mu}q'_{\alpha}\bar{u}_{\Lambda_c^+}(p',m'_s)\gamma_{\beta}u_p(p,m_s)}
{(q'-q)^2-m^2_{D^*}}F_{D^*}(q'-q)
\label{eq:lagDsexchange}\,.\,
\end{eqnarray}
Similarly, the $\bar{D}^{0}$-exchange amplitude for
$\gamma p \to \bar{D}^{*\,0} \Lambda^+_c$
is
\begin{eqnarray}
j^{\nu}_{D\,\,m'_s,\,m_s}(q,p,q',p')
&=&\frac{g_{D^*D\gamma}g_{ND\Lambda_c^+}}{\sqrt{m_{D^*}m_D}}
\epsilon^{\mu\nu\alpha\beta}
\frac{q_{\mu}q'_{\alpha}\epsilon_{D^*\,\beta}\bar{u}_{\Lambda_c^+}(p',m'_s)\gamma_{5}u_p(p,m_s)}
{(q'-q)^2-m^2_{D^*}}F_{D}(q'-q)
\label{eq:lagDsexchange}\,,\,
\end{eqnarray}

\subsection{$N^*_{c\bar{c}}$-excitation amplitudes}

The formula for calculating the
 resonant amplitude $\gamma N \rightarrow N^*_{c\bar{c}}\rightarrow \Lambda^+_c\bar{D}^*$
are the same as Eqs.(\ref{eq:famp-1})-(\ref{eq:famp-3}) except
 that the coupling constants $g_{2J\,J/\psi}$ for $J=1/2,3/2,5/2$ are
replaced by $g_{N^*\rightarrow \bar{D}\Lambda}$ for each $J$ listed  in Table \ref{tab:mwb2}.

For the $\gamma N \rightarrow N^*_{c\bar{c}}\rightarrow \Lambda^+_c\bar{D}$, we define
 $N^*_{c\bar{c}}\Lambda^+_c\bar{D}$ vertices  as follows:
\begin{eqnarray}
{\cal M}_{N^*(\frac{1}{2}^-)\Lambda^+_c \bar{D}}&=&
g_{1}\,\bar{u}_{\Lambda^+_c}u_{N^*}\label{eq:lagN12LD}\,,
\\
{\cal M}_{N^*(\frac{3}{2}^-)\Lambda^+_c \bar{D}}&=&
\frac{g_{3}}{m^2_D}\,\bar{u}_{\Lambda^+_c}\gamma_5\gamma_\mu u_{N^*\,\,\nu}
\,p^{\mu}_{\bar{D}}\,p^{\nu}_{\bar{D}}
\label{eq:lagN32LD}\,,\,
\\
{\cal M}_{N^*(\frac{5}{2}^+)\Lambda^+_c \bar{D}}&=&
\frac{g_{5}}{m^3_D}\,\bar{u}_{\Lambda^+_c}\gamma_5\gamma_\mu
u_{N^*\,\,\nu\lambda}\,p^{\mu}_{\bar{D}}\,p^{\nu}_{\bar{D}}\,p^{\lambda}_{\bar{D}}
\label{eq:lagN52LD}\,,
\end {eqnarray}
where $p_{\bar{D}}$ is the four momentum of $\bar{D}$ meson.
The coupling can be calculated from the partial decay widths
listed in No. 6, 9, and 18 of Tab.\ref{tab:mwb2}. We then get
 $g_1 = 0.40$, $g_3 = 1.29$, and $g_5 = 13.39$.

With the above equations and the $\gamma p \to N^*_{c\bar{c}}$ given in Table \ref{tab:coup-g}, we can
calculate the amplitude for $\gamma p \to N^*_{c\bar{c}} \to \bar{D}^0 \Lambda^+_c$ and
obtain the corresponding current matrix element   $j^{\nu}_{N^*\,\,m'_s,\,m_s}(q,p,q',p')$
( defined in Eq.(\ref{eq:tot-amp})) as:

\begin{eqnarray}
j_{N^*(\frac{1}{2}^-)\,\,m'_s,\,m_s}^{\nu}(q,p,q',p') &=&
\sum_{V=J/\psi,\rho,\omega}
g_{1}
\frac{\bar{u}_{\Lambda^+_c}(p',m'_s)
\left(\slashed{q}+\slashed{p}+m_{N^*_{c\bar{c}}}\right)
u_p(p,m_s)}
{W^2-m^2_{N^*_{c\bar{c}}}+i\Gamma_{N^*_{c\bar{c}}}m_{N^*_{c\bar{c}}}}
F_V(0)
\nonumber\\&&\times
\frac{ie}{f_{V}}
\frac{-m^2_{V}\tilde{g}_{1V}}
{-m^2_{V}+i\Gamma_{V}m_{V}}
\gamma_{5}\tilde{\gamma}_{\beta}
\left(
g^{\beta\nu}
-
\frac{3}{2}\frac{\tilde{r}^{\beta}\tilde{r}^{\nu}}{\tilde{r}^2}
+\frac{1}{2}\tilde{g}^{\beta\nu}_{N^*}\right),
\label{eq:famp-f}
\\
j_{N^*(\frac{3}{2}^-)\,\,m'_s,\,m_s}^{\nu}(q,p,q',p')  &=& \sum_{V=J/\psi,\rho,\omega}\frac{g_{3}}{m^2_D}
\frac{
\bar{u}_{\Lambda^+_c}(p',m'_s)
\gamma_5\slashed{q}'
(\slashed{q}+\slashed{p}+m_{N^*_{c\bar{c}}})P_{\alpha\beta}^{\frac{3}{2}}(p+q)
u_p(p,m_s)}
{W^2-m^2_{N^*_{c\bar{c}}}+i\Gamma_{N^*_{c\bar{c}}}m_{N^*_{c\bar{c}}}}
\nonumber\\&&\times
q'^\alpha F_V(0)\frac{ie}{f_{V}}
\frac{-m^2_{V}\tilde{g}_{3V}}
{-m^2_{V}+i\Gamma_{V}m_{V}}
\left(
g^{\beta\nu}
-\frac{3}{2}\frac{\tilde{r}^{\beta}\tilde{r}^{\nu}}{\tilde{r}^2}
+\frac{1}{2}\tilde{g}^{\beta\nu}_{N^*}\right),
\\
j_{N^*(\frac{5}{2}^+)\,\,m'_s,\,m_s}^{\nu}(q,p,q',p')&=&\sum_{V=J/\psi,\rho,\omega}
\frac{g_{5}}{m^3_D}
\frac{
\bar{u}_{\Lambda^+_c}(p',m'_s)\gamma_5\slashed{q}'
(\slashed{q}+\slashed{p}+m_{N^*_{c\bar{c}}})
P_{\alpha\beta\,\alpha'\beta'}^{\frac{5}{2}}(p+q)
u_p(p,m_s)}
{W^2-m^2_{N^*_{c\bar{c}}}+i\Gamma_{N^*_{c\bar{c}}}m_{N^*_{c\bar{c}}}}
\nonumber\\&&\times
q'^\alpha q'^\beta F_V(0)
\frac{ie}{f_{V}}
\frac{-m^2_{V}\tilde{g}_{5V}/m_N}
{-m^2_{V}+i\Gamma_{V}m_{V}}
\nonumber\\&&\times
\left(g^{\nu\alpha'}\tilde{r}^{\beta'}
-\frac{5}{3}\frac{\tilde{r}^{\nu}\tilde{r}^{\alpha'}\tilde{r}^{\beta'}}{\tilde{r}^2}
+\frac{1}{3}\left(
\tilde{g}^{\nu\alpha'}_{N^*}\tilde{r}^{\beta'}
+\tilde{g}^{\nu\beta'}_{N^*}\tilde{r}^{\alpha'}
+\tilde{g}^{\alpha'\beta'}_{N^*}\tilde{r}^{\nu}\right)
\right),
\end{eqnarray}

\subsection{Predictions of total cross sections}

\begin{figure}[htbp] \vspace{-0.cm}
\begin{center}
\includegraphics[width=1.0\columnwidth,angle=0]{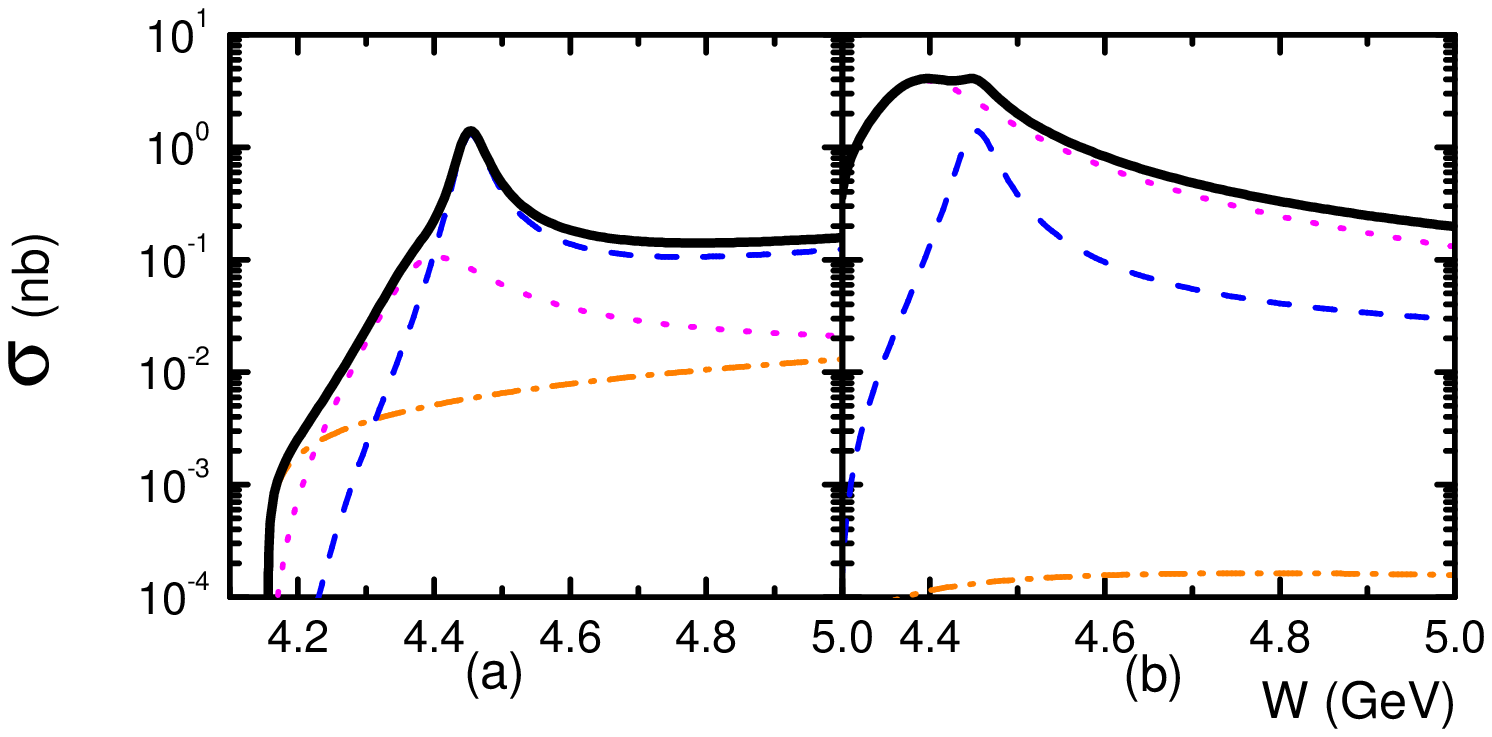}
\caption{The total cross sections of $\gamma p \to \bar{D}^0 \Lambda^+_c$(a) and $\gamma p \to \bar{D}^{*0} \Lambda^+_c$(b) with the invariant mass of $\gamma p$.
The pink dotted, blue dashed, orange dotted-dashed lines are the contribution purely from the $N^*_{c\bar{c}(\frac{3}{2}^-)}$, $N^*_{c\bar{c}(\frac{5}{2}^+)}$, and the background with $\Lambda=0.55$ GeV, respectively.
The black solid thick lines are for the coherent summation of all contributions with background $\Lambda=0.55$ GeV.
}
\label{fig:ndlc}
\end{center}
\end{figure}

The predicted total cross section of $\gamma p \to N^*_{c\bar{c}} \to \bar{D}^0 (\bar{D}^{*\,0}) \Lambda^+_c $ are shown in Fig.\ref{fig:ndlc}.
All calculations are done with cutoff $\Lambda=550$ MeV, as determined in the previous sections for $J/\psi$ production.
We first find that the meson-exchange contributions (dotted-dotted-dashed) to the predicted total cross section of $\gamma p \to N^*_{c\bar{c}} \to \bar{D}^0 (\bar{D}^{*\,0}) \Lambda^+_c $ are very weak.
%
%
The contribution from $N^*_{c\bar{c}}(\frac{3}{2}^-)$(blue dashed) is larger than that of $N^*_{c\bar{c}}(\frac{5}{2}^-)$( orange dotted).
Clearly, if the predicted cross section given in
 Fig.\ref{fig:ndlc} can be measured, it will provide an additional
test of the prediction of $N^*_{c\bar{c}}(\frac{3}{2}^-)$ state.
Hopefully such measurements can be made in the near future
as an additional test of our prediction on $J/\psi$ production, presented in the previous section.

\begin{table}[b]
      \setlength{\tabcolsep}{0.15cm}
 \begin{center}\caption{The coupling of  $g_{N^*\to\bar{D}\Lambda_c}$ and $g_{\bar{D}^{*}\Lambda_c}$ are used in the calculation.}
 \begin{tabular}{ccccccccccc}\hline
No. &    $J^{P}$ & $m$  & $\Gamma_{\bar{D} \Lambda_c}$ & $g_{N^*\to\bar{D}\Lambda_c}$  & $\Gamma_{\bar{D}^* \Lambda_c}$
 & $g_{N^*\to\bar{D^*}\Lambda_c}$ & Ref.     \\
    \hline
6 &  $\frac{1}{2}^{-}$  & $4481$  & $1.02$ & $0.40$   & $0.3$   & $0.043$
  & \cite{Xiao:2013yca}       \\
9 &  $\frac{3}{2}^{-}$  & $4380$  & $1.2$  & $1.29$   & $131.3$ & $1.90$
  & \cite{Lin:2017mtz}       \\
18 &  $\frac{5}{2}^{+}$  & $4450$  & $18.8$ & $13.39$  & $20.5$  & $2.18$
  & \cite{Lin:2017mtz}       \\
\hline
 \end{tabular}  \label{tab:mwb2}
 \end{center}
\end{table}

\begin{figure}[htbp] \vspace{-0.cm}
\begin{center}
\includegraphics[width=0.25\columnwidth,angle=270]{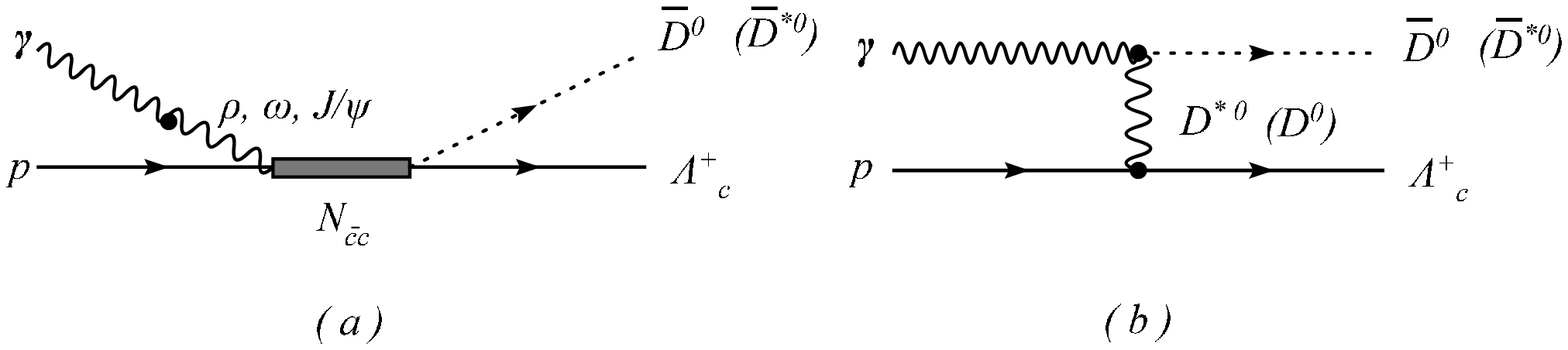}
\caption{The diagram (a) for $\gamma p \to N_{c\bar{c}} \to  \bar{D}^{0}(\bar{D}^{*0}) \Lambda^+_c $ with the VMD by $\rho$, $\omega$ and $J/\psi$ coupled with $\gamma$, and (b) for $\gamma p  \to  \bar{D}^{0} (\bar{D}^{*0}) \Lambda^+_c $  by exchanging $D^0 (D^{*0})$.
}
\label{fg:ndslc}
\end{center}
\end{figure}

%
%
%


\section{Summary}
By using the predictions from the available
meson-baryon  coupled-channel models, we have investigated
the excitations of nucleon resonances with hidden charm, $N^*_{c\bar{c}}$,
in the $\gamma\,\, p$ reactions.
For the $\gamma\,p\rightarrow J/\psi\, p$ process,
 the Pomeron-exchange model
of Donnachie and Landshoff,
  with the
parameters determined from
fitting the available  total cross section data up to $W=300$ GeV, is used to calculate the non-resonant
amplitudes.
The resonant $\gamma p \rightarrow N^*_{c\bar{c}}\rightarrow J/\psi\, p$ amplitudes are calculated by
using (1) the partial decay widths predicted by the
considered meson-baryons coupled-channel models  to evaluate the
$N^*_{c\bar{c}}\to MB$ transition matrix elements, and (2) the Vector Meson Dominance (VMD) model
to evaluate $\gamma p \to N^*_{c\bar{c}}$ as  $\gamma p \to V p \to N^*_{c\bar{c}}$ with V = $\rho$,
$\omega$, $J/\psi$.
The predictions from adding these two amplitudes
then depend on an off-shell form factor
$F_V(q^2)=\lambda^4/(\Lambda^4 + (q^2 - m^2_V)^2)$ which is needed to account for the $q^2$-dependence of VMD model.
We find   that with $\Lambda = 0.55$ GeV, the predicted total cross sections of
 $\gamma\,p\rightarrow J/\psi\, p$
 are within
the range of the available data in the energy region near $J/\psi$ production threshold.
We then demonstrate that the $N^*_{c\bar{c}}$ can be most easily identified in the differential cross sections at large angles where the contribution from Pomeron-exchange becomes negligible.

With the same VMD model and the same coupled-channel model of $N^*_{c\bar{c}}$,
we then predict the cross sections of
 $\gamma\,p\rightarrow \rightarrow \bar{D}^0\Lambda_c^+ (\bar{D}^{*0}\Lambda_c^+)$.
We suggest that experiments on these reactions can be more effective to study $N^*_{c\bar{c}}$
since their non-resonant amplitudes,
due to the exchange of $\bar{D}^{*0}$ ($\bar{D}^{0})$, are found to be very weak.

The most unsatisfactory aspect of this work is the phenomenological determination of
the off-shell form factor $F_V(q^2)$. It is determined by only using the data of total cross sections
of $\gamma\,p\rightarrow J/\psi\, p$ near the threshold, shown in Fig.\ref{fg:backg}.
While our predictions could be used  as a first-step to determine whether the $N^*_{c\bar{c}}$
predicted by the available meson-baryon coupled-channel models can be found in the new data from JLab,
it is necessary to develop a more fundamental approach to also predict $F_V(q^2)$ from QCD  models.
Obviously, such an improvement is necessary for using the $q^2$-dependence of the
$J/\psi $ electro-production cross section data to investigate nucleon resonances with hidden charm.
\clearpage

\begin{acknowledgments}
One of the authors Jia-jun Wu want to thanks the useful discussion with Hai-qing Zhou, Jun He, Jujun Xie, Qiang Zhao.
This project is supported by the Thousand Talents Plan for Young Professionals and the National Natural Science Foundation of China under Grants No. ~11621131001 (CRC110 cofunded by DFG and NSFC) and Grant No.~11835015.
This work is also partially supported
by the U.S. Department of Energy, Office of Science, Office of Nuclear Physics, Contract No. DE-AC02-06CH11357.

\end{acknowledgments}

\bibliography{refs}

\end{document}